\documentstyle[12pt]{article}
\def\be{\begin{eqnarray}}
\def\ee{\end{eqnarray}}
\def\ba{\begin{array}}
\def\ea{\end{array}}

\begin{document}

\begin{center}
{\bf\LARGE {Arrow of time in generalized quantum theory \\
\vskip 3mm and its classical limit dynamics}}

\end{center}

\vskip 1cm
\begin{center}
{\bf \large {Vadim V. Asadov$^{\star}$\footnote{asadov@neurok.ru}
and Oleg V.
Kechkin$^{+\,\star}$}\footnote{kechkin@depni.sinp.msu.ru}}
\end{center}

\vskip 5mm

\begin{center}
$^+$Institute of Nuclear Physics,\\
Lomonosov Moscow State University, \\
Vorob'jovy Gory, 119899 Moscow, Russia
\end{center}

\vskip 5mm

\begin{center}
$^\star$IIP--Neur\,OK,\\
Scientific park of MSU, Center for Informational Technologies--104,\\
Vorob'jovy Gory, 119899 Moscow, Russia
\end{center}

\vskip 1cm

PACS No(s).\, : 05.30.-d,\,\,05.70.-a.



\renewcommand{\theequation}{\thesubsection.\arabic{equation}}

 \vskip 1cm
\begin{abstract}
In this paper we have studied a generalized quantum theory and its
consistent classical limit, which possess a well-defined arrow of
time in their dynamics. The original quantum theory is defined as
analytically dependent on complex time and specified by
non-Hermitian Hamiltonian structure.
\end{abstract}




\section{Generalized quantum theory}

Irreversibility of evolution is a common feature of all real
dynamical systems. This fact is reflected in the second law of
thermodynamics, which states that entropy of any closed system can
not decrease. Figuratively, one says about some `arrow of time',
which separates past and future in the absolute manner, and must be
guaranteed by any realistic dynamical theory without fail
\cite{atf}--\cite{atl}.

Here it is important to stress, that thermodynamics itself is not a
dynamical theory of the same type as, for example, classical or
quantum mechanics. Rather, it provides some general theoretical
framework for dynamical theories of this type, which pretend to the
adequate description of evolving physical reality. Also, it can be
said that thermodynamical principles must be realized on the base of
the consistent dynamical theory of the fundamental type, and that
thermodynamics imposes hard restrictions to corresponding
theoretical constructions \cite{td}.

In this connection, one must take into account that all quantum and
classical fundamental theories are conservative Hamiltonian systems
\cite{qm}--\cite{cm}. All these systems are reversible in time,
because one can interchange the initial and final data to obtain
really possible result of their dynamical evolution. Thus, these
theories, being the most natural concrete classes of the closed
dynamical systems, can not be used for any consistent realization of
the thermodynamic conception of the arrow of time. Actually, all
attempts performed to `average out' the reversible results of these
theories to obtain the thermodynamic irreversible ones, contain
hidden incorrect actions.

Also, impossibility to consider unidirectional evolution in
framework of the classical mechanics can be understood using results
of the Poincare theorem. Actually, it states, that one can decompose
any classical motion to the set of the Poincare cycles, so this
motion obtains transparently reversible character. The same
situation takes place in the quantum mechanics: using the
Hamiltonian proper basis one reduces quantum motion to the set of
corresponding oscillations, which are reversible manifestly. As the
result, the probabilities to find quantum system in the Hamiltonian
eigenstates are constant quantities, and one can not relate any
irreversible dynamics of the kinetic type to them.

In this point we would like to emphasize, that the set of
conventional kinetic equations contain arrow of time by its
definition. Such equations had been used extensively by Prigogine to
study both the thermodynamic and synergetic processes \cite{Pr}. The
only conceptual problem of the all corresponding approaches is how
to ground the kinetic method of realization of the time irreversible
disciplines on the fundamental level in the really correct form. In
the over words, one must show how to modify the fundamental theories
of the classical and quantum mechanic types to achieve their
consistency with conventional kinetic framework. In this paper we
answer on this question by presentation of the corresponding
generalizations of the both quantum and classical theories.

Namely, we consider some natural generalization of the conservative
quantum theory, and show that its dynamics is actually irreversible.
Our generalization is related to use of the both complex parameter
of evolution (or `complex time', \cite{ctf}--\cite{ctl}) and
non-Hermitian Hamiltonian independent on this evolutionary parameter
\cite{NonHerm-f}--\cite{NonHerm-l}.  Also, we suppose analytic
dependence of quantum state vectors on the evolutionary parameter
and impose commutativity restriction on the Hamiltonian and the
result of its Hermitian conjugation. We interpret real and imaginary
parts of the complex time as `usual' time and minus half of inverse
absolute temperature, respectively. Then, we identify Hermitian part
of the Hamiltonian with energy operator of the quantum system,
whereas anti-Hermitian one we relate to operator of decay times for
the energy eigenstates. Finally, we define thermodynamic regimes of
evolution of the quantum system; this allows us to determine
temperature functions for concrete thermodynamic processes
considered in the paper. All these regimes are obtained by help of
fixation of average values of the corresponding observable
quantities and have natural and physically well-motivated sense.

The work is organized as follows. It consists of two parts, which
describe the quantum and classical modified theories, both
Hamiltonian ones and time-irreversible. The classical theory is
derived as the corresponding limit of the quantum theory case; we
study it under the same conditions as its fundamental quantum
origin.

First of all, we present general scheme for modification of the
quantum theory supplemented by method of fixation of the
thermodynamic regime. After that, we study isothermal and adiabatic
regimes in details and show, that both these types of evolution
actually possess the arrow of time. It is proved, that effect of the
irreversibility can be related to the non-increasing evolution of
the average value of the anti-Hermitian part of the Hamiltonian of
the system in these cases.

Then, we investigate dynamics of quantum subsystem of the
generalized quantum system with two different values of the decay
time parameter and study the entropy evolution of this subsystem. In
particular, we show how one can guarantee existence of the
well-defined arrow of time in this dynamics by some additional
hipping of the decay properties of the vacuum state only.

Also, we study the non-Hermitian generalization of the quantum
theory with discrete symmetry of the parity type and show, that
physical evolution in this theory possesses well-defined arrow of
time. The result of its evolution is related to transformation of
the `left' quantum world to the `right' one (or vice versa) for the
arbitrary thermodynamic regime. We have realized this general
quantum scheme for massless Dirac particles to explain the problem
of helicity symmetry breaking for the real neutrino in the natural
framework of the modified quantum dynamics.

In the second part of the work we derive a modified Hamiltonian
dynamics, which describes a classical regime of evolution of the
quantum theory studied in the first part. We define the classical
dynamics as such special limit of the quantum one, that is specified
by resonance character of the probability density in the theory
representation taken. In the `hard'\, classical mode of the theory
we put $\hbar\rightarrow 0$; in the `soft'\, regime of its evolution
we save the Planck constant as some free parameter.

Our starting point is the representation of the original quantum
theory in terms of the quantum numbers, which contain the energy and
decay operator indexes. We consider the field of `big values'\, of
the all quantum numbers and replace the originally discrete problem
to the continuous one in the appropriate way. After that, we check
out the consistency of the equations, which describe the
corresponding `hard'\, dynamics. Note, that this point is really
nontrivial one, because this dynamics depend on the two evolutionary
parameters: the physical time and the temperature. Thus, the
`hard'\, classical dynamics under discussion is, in fact, some
special classical field theory, which is described by some system of
partial differential equations. We show, that this classical
dynamics possesses arrow of time like its time-irreversible quantum
origin (in the isothermal and adiabatic regimes of evolution, at
least).

Then, we establish a symplectic group of hidden symmetries for the
leading term of the modified classical Hamiltonian equations for the
`soft'\, dynamics with $\hbar\neq 0$. Namely, we have shown, that in
this case one deals with the symmetry group $Sp(2N,\,R)$ for the
theory with $N$ degrees of freedom. It is shown, that this symmetry
corresponds to some subgroup of the group of conventional  canonical
transformations. The most interesting opportunity is related to fact
that this hidden symmetry  acts not only on the `usual'\,
Hamiltonian part of the equations, but it also preserves the main
additional term, related nontrivially to the original quantum
theory. This perturbation-like term is defined by the classical
limit of the dispersion characteristics of the quantum theory; it
represent the classical dynamics as some effective `history of of
the world tubes'.\, The `world tubes'\, arise in the theory under
consideration instead of origination of the world lines in the
standard classical mechanics. Our effective `fat'\, phase
trajectories seem really adequate alternative of the `thin'\, ones
if one deals not only with the average quantities, but with their
dispersions too.

We have studied in details the symplectic symmetry structure of the
generalized classical theory. It is shown, that this structure
possesses exactly the matrix-valued General Relativity form. For
example, we have found the nonlinear sector of these symmetries and
identified it as the matrix-valued Ehlers transformation from the
standard Einstein theory. We think, that these established
continuous Lee symmetries of the theory can help in construction of
the solution of the main dynamical problem, i.e. in integration of
the motion equations. It seems possible, that one can obtain a
constructive approach to study in details the time irreversible
classical dynamics supplemented by the fundamental quantum theory of
the type, presented in this work.

\subsection{Quantum dynamics and thermodynamics}

\setcounter{equation}{0}

It seems clear, that any realistic generalization of the quantum
theory must start with some complex linear space of state vectors
$\Psi_{1},\,\,\Psi_{2}$, etc., with some well-defined scalar
products $\Psi_{1}^+\Psi_{2}$, etc., of these vectors. Moreover, one
must preserve a probability interpretation of the theory. For
example, in the normalizable case one must use the relation
\be\label{G1}
P=\frac{\Psi_{1}^+\Psi_{2}\,\cdot\,\Psi_{2}^+\Psi_{1}}{\Psi_{1}^+\Psi_{1}\,
\cdot\,\Psi_{2}^+\Psi_{2}} \ee for the probability $P$ to find the
system in the condition with the state vector $\Psi_{1}$ when this
system is specified by the state vector $\Psi_{2}$. Then, it is
necessary to deal with some set of the observables $Q_1,\,Q_2$,
etc., which are linear Hermitian operators acting on the space of
the state vectors. Without any doubt, the rule for calculation of
the average value $\bar Q$ for the observable $Q$, which is related
to the state vector $\Psi$, must save its well-known conventional
form: \be\label{G2} \bar{Q}=\frac{\Psi^+{{Q}}\Psi}{\Psi^+\Psi}. \ee
Also, the average values of the observable quantities must preserve
their meaning in relation of the results of the quantum theory to
the corresponding observations in the real world.

Then, dynamical aspects of the generalized quantum theory must be
described by the help of some evolutionary parameter $\tau$ and
Hamiltonian operator ${\cal{H}}$. For conservative systems one has
${\cal{H}}_{,\tau}=0$ and $\Psi=\Psi (\tau)$ in the
Schr\"{o}dinger's picture which we use in this work. The main
dynamical equation of the theory also must preserve its conventional
Schr\"{o}dinger's form, \be\label{G3} i\hbar
\Psi_{,\tau}={\cal{H}}\Psi. \ee Note, that all elements of the
quantum theory listed above are completely standard ones.

Our modification of the quantum theory is based on the use of the
complex time parameter $\tau\neq\tau^*$ and non-Hermitian
Hamiltonian $\cal{H}\neq\cal{H}^+$. We consider a holomorphic
variant of the theory, with $\Psi_{,\tau^{*}}=0$ and
${\cal{H}}_{,\tau^{*}}=0$, which leads to the simplest
generalization of the standard quantum theory. We restrict our
consideration by the theories with $\left [
\cal{H},\,\,\cal{H}^+\right ]=0$; this last relation seems the most
natural continuation of the one for the standard theory. Actually,
it becomes an identity when ${\cal{H}}={\cal{H}}^+$. Then, we
parameterize the complex evolutionary parameter $\tau$ in terms of
the real variables $t$ and $\beta$ in the following way:
\be\label{G5} \tau=t-\,i\,\frac{\hbar}{2}\,\beta, \ee whereas the
non-Hermitian operator ${\cal{H}}$ will be represented using the
Hermitian ones $E$ and $\Gamma$ as \be\label{G6}
{\cal{H}}=E-\,i\,\frac{\hbar}{2}\,\Gamma. \ee Note, that
\be\label{G4'}[E,\,\,\Gamma]=0, \ee in accordance to the restriction
imposed above. In the next section we will argue that if one relates
the quantities $t$ and $E$ to the `usual' time and energy operator,
respectively, then the remaining ones $\beta$ and $\Gamma$ mean the
inverse temperature $\beta=1/T$ and the operator of decay parameters
of the quantum system. Also, it will be demonstrated that the last
operator defines the arrow of time, which seems naturally in view of
the transparent irreversibility of the any decay process.

Then, this scheme of generalization of the quantum theory must be
completed by introducing of a conception of the thermodynamic
regime, which has a form of fixation of the temperature function,
i.e. of the relation $\beta=\beta (t)$. For example, one can study
evolution of the system in the isothermal case $\beta=\rm{const}$;
also it is possible to analyze the adiabatic situation with
$\bar{E}(t,\beta)=\rm{const}$. The most general thermodynamic
regime, which is related to some observable quantity $Q$, can be
defined in the following form: \be\label{G7} f\left ( t,\,
\beta,\,\bar{{Q}}(t,\beta\right )=0. \ee Here it is important to
stress, that our main goal is related to study of the time arrow
effect for class of the quantum theories defined by the relations
(\ref{G1})--(\ref{G6}) and the restriction (\ref{G7}). Below it is
shown that this effect actually takes place in the physically
well-motivated thermodynamic regimes. Thus, the general theoretical
scheme given above provides really natural tool for incorporating of
the Second Law of thermodynamics into the pure Hamiltonian quantum
theory. Note, that this approach leads not only to the unification
of thermodynamics with quantum theory. Also, the unified theory of
thermodynamics with classical mechanics (or with classical field
theory) can be obtained in the classical limit of the corresponding
quantum unification scheme.


\subsection{Quantum thermodynamic regimes}

\setcounter{equation}{0}

So, our main goal is to detect and to study the irreversible aspects
of evolution of the generalized quantum system defined in the
previous section. To do it, let us express the significant
quantities of the theory in terms of the basis of eigenvectors
$\psi_n$ of the commuting operators $E$ and $\Gamma$. We take it in
its orthonormal form, i.e. we mean that the identities
$\psi_n^+\psi_k=\delta_{nk}$ take place for all values of the
indexes $n$ and $k$. Here, of course, these indexes must be
understood in the appropriate multi-index sense, and all the
summations arising below are of the corresponding generalized type.

The eigenvalue problem under consideration reads: \be\label{S1}
E\psi_n=E_n\psi_n, \qquad \Gamma\psi_n=\Gamma_n\psi_n; \ee it can be
reformulated in terms of the non-Hermitian operator ${\cal H}$.
Actually, it is easy to see, that $\psi_n$ is the eigenvector for
this operator, which corresponds to the complex eigenvalue ${\cal
H}_{n}=E_n-i\hbar/2\,\Gamma_n$. Then, the state vectors
$\Psi_n=e^{-i{\cal H}_{n}\tau/\hbar}\psi_n$ satisfy the
Schr\"{o}dinger's equation (\ref{G3}), and also form the complete
(but $\tau$-dependent) basis. This basis can be used for
representation of any solution $\Psi$ of the Schr\"{o}dinger's
equation in the form of linear combination with some set of constant
parameters $C_n$, i.e., as \be\Psi=\sum_n C_n\Psi_n.\ee Using this
decomposition and the basis properties, let us calculate the
probability $P_n$ to find the quantum system in its basic state
$\Psi_n$, when it is described by the state vector $\Psi$. After the
application of Eq. (\ref{G1}) one obtains, that \be\label{S2}
P_n=\frac{w_n}{Z}, \ee where $Z=\sum_n w_n$, \be\label{S3}
w_n=\rho_n e^{-(E_n\beta+\Gamma_nt)}, \ee and $\rho_n=|C_n|^2$.
Note, that all the following quantum analysis will be related to the
study of dynamics of the probabilities (\ref{S2})--(\ref{S3}) in the
different thermodynamical regimes and for the different special
realizations of the generalized quantum theory. In this analysis,
the formula (\ref{S3}) provides the base for the both theoretical
and experimental study of the our generalization comparing with the
standard quantum theory.

First of all, let us consider the evolution of two very special
systems which will explain our interpretation of the imaginary part
of the complex parameter $\tau$ and the anti-Hermitian one of the
Hamiltonian $\cal H$. Namely, the first system has the coinciding
eigenvalues $\Gamma_n$ for the all indexes $n$. It is easy to see,
that $w_n=\rho_ne^{-E_n\beta}$ in this case. This means, that the
quantity $\beta$ is actually the inverse absolute temperature if
$E_n$ had been identified as the $n$-th energy level of the system.
The second special system will be specified by the coinciding
eigenvalues $E_n$. Then $w_n=\rho_ne^{-\Gamma_nt}$, so the
quantities $\Gamma_n$ have the sense of the decay parameters, if $t$
means the conventional (`usual') time. Actually, let us consider,
for example, the system with $\Gamma_{n_{\star}}=\min_n
\{\Gamma_n\}=0$ in the situation where the single level $n$ is
weakly excited under the level $n_{\star}$. In this special case
$\rho_{m}=\rho_{m}\delta_{mn}$, where $n,m \neq n_{\star}$, and also
$\rho_{n_{\star}}\approx 1$ whereas $\rho_{m}\approx 0$. It is easy
to see, that for this quantum state
$P_{m}\approx\rho_{m}e^{-\Gamma_{m}t}$, so $t_{m}=1/\Gamma_{m}$ is a
conventional `time of life' of the basis state which has the energy
$E_m$.

Note, that in the same situation with $\Gamma_{n_{\star}}=\max_n
\{\Gamma_n\}$, one deals with the exponentially increasing
probability $P_{m}(t)$. However, we use the term `parameters of
decay' for the quantities $\Gamma_n$ for all regimes of the
evolution. Then, it is clear that the solution space of the theory
of the discussing type can be decomposed into the direct sum of the
solution subspaces which have a given value of the energy or of the
parameter of decay. For all these subspaces the interpretation of
$\beta$ and $\Gamma_n$ is the same as for the special systems of the
first and of the second types discussed above, respectively. In
fact, we extend the interpretation of these physical quantities (as
well as the interpretation for the quantities E and t) to the total
solution space of the generalized quantum theory, making a simple
and really natural hypothesis.

Now let us consider the non-specified quantum theory of the form,
presented in the previous section, and fix the isothermal regime of
its thermodynamical evolution. It is easy to prove, that in the case
of $\beta=\rm{const}$, the dynamical equations for the basic
probabilities read: \be\label{S4} \frac{dP_n}{dt}=-\left (
\Gamma_n-\bar{\Gamma}\right )P_n .\ee To perform the analysis, let
us study a behavior of the quantity $\bar{\Gamma}$. After some
calculations one obtains, that \be\label{S5}
\frac{d\bar{\Gamma}}{dt}=-D_{\Gamma}^2,\ee where
$D_{\Gamma}^2=\overline {\left ( \Gamma-\bar{\Gamma}\right )^2}$ is
the squared dispersion of the quantum observable $\Gamma$. From Eq.
(\ref{S5}) it follows, that the function $\bar\Gamma (t)$ is not
increasing, so the isothermal regime actually allows the time arrow
which can be naturally related to the average value of the decay
operator $\Gamma$.

Then, the probability $P_n$ rises when $\bar{\Gamma}>\Gamma_n$ and
degenerates if $\bar{\Gamma}<\Gamma_n$. Thus, in the isothermal
regime the quantity $|\Gamma_n-\bar{\Gamma}|$ has a sense of
logarithmic decrement of the growth or degeneration of the
probability to find the quantum system on the energy level $E_n$.
Also, in this regime one obtains the following picture for
asymptotics of the probabilities at $t\rightarrow +\infty$: all the
`activated'\, (i.e., with $\rho_n\neq 0$) probabilities with the
maximal value of the decay parameters rise droningly, whereas all
the remaining probabilities fall to the zero ones. Namely, let us
define the multi-index $n_{_\star}$ by the relation
$\Gamma_{n_{_\star}}=\min_n \{ \Gamma_{n}\}$, again.

Then, for the only non-degenerating probabilities $P_{n_{\star}}(t)$
one obtains the following asymptotical result: \be\label{S6}
P_{n_{_\star}}(+\infty)=\Pi P_{n_{_\star}}(0), \ee where the scale
parameter $\Pi>1$ reads: \be\label{S7} \Pi=1+\frac{\sum_{n\neq
n_{_\star}}\rho_ne^{-E_n\beta}}
{\sum_{n_{_\star}}\rho_{n_{_\star}}e^{-E_{n_{_\star}}\beta}}. \ee
Note, that the relations (\ref{S6})--(\ref{S7}) have the form of a
`dressing procedure'\, in the standard quantum field theory. This
circumstance seems really hopeful for renormalization of the quantum
field theories with non-Hermitian Hamiltonian and complex time
parameter. Actually, in the standard quantum field theory one deals
with the oscillating harmonics (in the corresponding representation
on shell), which cannot be eliminated without fail of the standard
mathematical logic. However, one needs in cut of the increasing
modes to reach the theoretical scheme with finite calculations. It
is a well known fact, that all known cut procedures are in
contradiction with all `normal neglecting principles'.\, The
generalized quantum theory under consideration allows one to speak
about the modes which degenerate dynamically, and also about the
modes which remain `alive'\, at the `big times'.\, Moreover, these
last modes of the exact quantum theory solutions become
renormalized, if one compares their initial and final probabilities,
see Eqs. (\ref{S6})--(\ref{S7}).

Then, it is easy to prove, that in the general case of $\beta=\beta
(t)$, the dynamical equations for the basic probabilities have the
following form: \be\label{S8} \frac{dP_n}{dt}=-\left [
\Gamma_n-\bar{\Gamma}+\left ( E_n-\bar{E}\right
)\frac{d\beta}{dt}\right ] P_n. \ee Here the specific function
$d\beta/dt$ must be extracted from the corresponding thermodynamical
regime (\ref{G7}). In the adiabatic case, when
$\bar{E}=\sum_nE_nP_n=\rm{const}$, one obtains immediately that
\be\label{S9} \frac{d\beta}{dt}=-\frac{\overline{E\Gamma}-\bar
E\bar\Gamma}{D_{E}^2}, \ee where $D_E$ denotes the dispersion of the
energy operator $E$. Using Eqs. (\ref{S8})--(\ref{S9}), for the
dynamics of the quantity $\bar{\Gamma}$ in the adiabatic regime one
obtains: \be\label{S10} \frac{d\bar{\Gamma}}{dt}=-D_{\Gamma}^2 \left
[ 1- \frac{\left ( \overline{E\Gamma}-\bar E\bar \Gamma\right
)^2}{D_E^2D_{\Gamma}^2}\right ].\ee

Our goal is to show that the function $\bar{\Gamma}(t)$ is
non-increasing again, so the generalized quantum system possesses
the well-defined arrow of time in its adiabatic regime of evolution
as well as in the isothermal case. To do it, let us prove that the
expression $[...]$ in Eq. (\ref{S10}) is not negative. Let us
introduce the formal vector quantities $X$ and $Y$ with the
components $X_n=E_n-\bar E$ and $Y_n=\Gamma_n-\bar \Gamma$,
respectively, and with scalar product $(XY)=\sum_nP_nX_nY_n$. It is
easy to see, that in terms of these quantities
$[...]=1-(XY)^2/[(XX)(YY)]$, so this expression is really
non-negative in view of the general Cauchy-Buniakowski inequality.

At the end of this section we would like to note, that the function
$\bar\Gamma (t)$ can not provide the arrow of time for the arbitrary
regime of thermodynamic evolution of the generalized quantum system.
Actually, this statement becomes transparent, if one considers the
specific regime with $\bar\Gamma=\rm{const}$. It is clear, that the
study of the corresponding dynamics must be performed in some
completely different terms. In fact, we prove its irreversible
character using the entropy function for one special class of the
generalized quantum systems in the next section.


\subsection{Example I: Thermodynamics of quantum subsystem}

\setcounter{equation}{0}

Let us consider the theory of the discussing type with
$n=(0,\,\nu)$, \, $\Gamma_0=\gamma\neq 0$, and $\Gamma_{\nu}=0$ for
all values of the collective index $\nu$. Our nearest goal is to
study the nontrivial thermodynamical regime $\bar\Gamma=\rm {const}$
of the evolution of this system in the time arrow framework.

First of all, we rewrite the basic probabilities of the theory,
i.e., the quantities \be\label{E1}
P_{0}&=&\frac{\rho_0}{\rho_0+e^{\gamma t}\tilde Z},
\nonumber\\
P_{\nu}&=&\frac{e^{\gamma t}\tilde w_{\nu}}{\rho_0+e^{\gamma
t}\tilde Z}, \ee where $\tilde Z=\sum_{\nu}\tilde w_{\nu}$, \,
$\tilde w_{\nu}=\rho_{\nu}e^{-\tilde E_{\nu}\beta}$ and $\tilde
E_{\nu}=E_{\nu}-E_{0}$, in terms of the average magnitude
$\bar\Gamma$. The result reads: \be\label{E2}
P_{0}=\frac{\bar\Gamma}{\gamma}, \qquad P_{\nu}=\left (
1-\frac{\bar\Gamma}{\gamma} \right )\tilde P_{\nu}, \ee where
$\tilde P_{\nu}=\tilde w_{\nu}/\tilde Z$. Thus, in the case of
$\bar\Gamma=\rm {const}$, one has $P_{0}=\rm {const}$ and
$P_{\nu}\sim\tilde P_{\nu}$, so the dynamics of the system under
consideration is defined by the dynamics of its effective subsystem
with the basic probabilities $\tilde P_{\nu}$ completely.

Note, that this subsystem, which is characterized by the energies
$\tilde E_{\nu}$, their average value $\bar{\tilde
E}=\sum_{\nu}\tilde P_{\nu}\tilde E_{\nu}$, and the squared
dispersion $D_{\tilde E}^2=\overline{\left (\tilde E-\bar{\tilde
E}\right )^2}$, is the conventional thermodynamic system. Its
evolution is given by the temperature regime $\beta=\beta (t)$,
which can be extracted from the corresponding macroscopic condition
($\bar\Gamma=\rm {const}$ in the situation under consideration). For
such systems the dynamical equations read: \be\label{E3} \tilde
P_{\nu,\,t}&=&\beta_{,t}\left (\bar {\tilde E}-\tilde
E_{\nu}\right )\tilde P_{\nu},\nonumber\\
\bar{\tilde E}_{,t}&=&-\beta_{,t}D_{\tilde E}^2, \ee and the arrow
of time can be related to the both functions $\beta (t)$ and
$\bar{\tilde E}(t)$. Moreover, one can introduce the entropy $\tilde
S=-\rm{ln}\,\tilde Z-\beta\bar{\tilde E}$, which growth is
proportional to the one for the average energy: \be\label{E4} \tilde
S_{,t}=-\beta\bar{\tilde E}_{,t}. \ee
 However, in this
chapter we use the quantity $\bar{\tilde E}(t)$ as the Lyapunov
function as the most natural tool in the analysis of the dynamics of
the system under consideration.

It is easy to prove, that in the case of $\bar\Gamma=\rm{const}$,
the temperature regime is defined by the relation \be\label{E5}
t=\frac{1}{\gamma}\left \{\rm{ln}\,\left [\rho_0\left
(\frac{\gamma}{\bar\Gamma}-1\right )\right ]-\rm{ln}\,\tilde
Z\right\} \ee (note, that $\tilde Z=\tilde Z(\beta)$ here). From
this relation and Eq. (\ref{E3}) it follows, that \be\label{E6}
\bar{\tilde E}_{,t}=-\frac{\gamma}{\bar{\tilde E}}D_{\tilde E}^2,
\ee so the function $\bar{\tilde E}(t)$ defines the arrow of time in
the all regions of conservation of its sign.

In view of this fact, it is natural to restrict our consideration by
the systems which possess minimal or maximal value of their energy
spectrum. Namely, it is clear, that $\bar{\tilde E}>0$ if $E_0=\rm
{min}_n\left \{ E_n\right \}$ (the situation A), and $\bar{\tilde
E}<0$ if $E_0=\rm {max}_n\left \{ E_n\right \}$ (the situation B).
Also, let us call the theories with $\gamma<0$ and $\gamma>0$ as the
cases I and II, respectively. Then, the class of theories under
consideration splits into the four subclasses IA, IB, IIA and IIB --
in accordance with the sign of their decay parameter $\gamma$, and
with the `min/max' character of the energy level $E_0$. Finally, it
is seen that the theories IA and IIA are characterized by increasing
of their average energy $\bar{\tilde E}$, whereas the theories IB
and IIB correspond to the decreasing behavior of this quantity. It
is useful to note, that $|\bar{\tilde E}|$ is the Lyapunov function
for the theory under consideration for the all types of the spectrum
$\{E_n\}$ in the thermodynamical regime $\bar\Gamma=\rm{const}$, as
it follows immediately from Eq. (\ref{E6}). However, the quantity
$\bar {\tilde E}$ seems more physically motivated function in view
of its close relation to the entropy, see Eq. (\ref{E4}).

Now let us study the adiabatic regime of evolution of the quantum
system under consideration. First of all, we rewrite the basic
probabilities (\ref{E1}) in terms of the quantity $\bar E$, i.e.
using the average energy of the system. The result reads:
\be\label{E7} P_0=1-\frac{\bar{\cal E}}{\bar{\tilde E}}, \qquad
P_{\nu}=\frac{\bar{\cal E}}{\bar{\tilde E}}\tilde P_{\nu}, \ee where
$\bar{\cal E}=\bar E-E_0$. Thus, the dynamics of the total system is
defined by the dynamics of its effective subsystem (with the
probabilities $\tilde P_{\nu}$) again. From Eq. (\ref{E7}) it
follows, that \be\label{E8} 0\leq\frac{\bar{\cal E}}{\bar{\tilde
E}}\leq 1; \ee this double inequality defines the region of possible
values of the temperature parameter $\beta$ after imposing of the
condition $\bar E=\rm{const}$. The temperature regime in this case
is given by the relation \be\label{E9} t=\frac{1}{\gamma} \rm{ln}
\left [ \frac{\rho_0\bar{\cal E}}{\tilde Z \left ( \bar{\tilde
E}-\bar{\cal E} \right )}\right ]. \ee It defines the inverse
temperature evolution from the initial value $\beta_0=\beta (0)$,
where \be\label{E10} \bar{\tilde E}(\beta_0)=\bar{{\cal E}}\left
(1+\frac{\rho_0}{\tilde Z(\beta_0)}\right ), \ee to the final one
$\beta_{\rm{as}}$, where \be\label{E11} \bar{\tilde
E}(\beta_{\rm{as}})=\bar{{\cal E}}. \ee From Eq. (\ref{E9}) it
follows that the system actually achieves the final value of its
temperature at the time asymptotics $t\rightarrow +\infty$; this
opportunity explains the notations used.

Then, taking into account Eqs. (\ref{E6}) and (\ref{E9}), it is easy
to prove that \be\label{E12} \bar{\tilde E}_{,t}=-\gamma \left
(\bar{\tilde E}-\bar{\cal E}\right )\frac{D_{\tilde
E}^2}{\overline{\tilde{E}^2}-\bar{\tilde E}\bar{\cal E}}. \ee From
the inequality (\ref{E8}) it follows, that
$\overline{\tilde{E}^2}-\bar{\tilde E}\bar{\cal E}\geq D_{\tilde
E}^2$, so the sign of $\bar{\tilde E}_{,t}$ is opposite to the sign
of the quantity $\gamma \left (\bar{\tilde E}-\bar{\cal E}\right )$.
Thus, it is natural to consider two regions $\bar{\tilde
E}>\bar{\cal E}>0$ and $\bar{\tilde E}<\bar{\cal E}<0$ of the
possible temperature evolution, which are two branches of the region
defined by inequality (\ref{E8}). Then, these two regions correspond
exactly to the situations A and B for the thermodynamical regime
$\bar\Gamma=\rm{const}$ studied above. Using this opportunity, we
can realize these two situations by consideration of the theories
with $E_0=\rm{min}\left\{ E_n\right\}$ and $E_0=\rm{max}\left\{
E_n\right\}$, respectively, again. Finally, it is easy to see, that
behavior of the Lyapunov function $\bar{\tilde E}(t)$ in the both
thermodynamical regimes ($\bar\Gamma=\rm{const}$ and $\bar
E=\rm{const}$) is the same for the all four variants of the theory
(IA, IB, IIA and IIB).

At the end of the analysis of the adiabatic regime we would like to
stress, that the quantity $\bar\Gamma$ is the decreasing Lyapunov
function for all signs of the decay parameter $\gamma$ and for all
types of the energy spectrum $E_n$. Namely, one can check, that
\be\label{E13}\bar{\Gamma}_{,t}=-\gamma^2\frac{\bar{\cal
E}}{\bar{\tilde E}}\left (1-\frac{\bar{\cal E}}{\bar{\tilde
E}}\right )\frac{D_{\tilde E}^2}{\overline{\tilde{E^2}}-\bar{\tilde
E}\bar{\cal E}};\ee so $\bar{\Gamma}_{,t}\leq
-\gamma^2\frac{\bar{\cal E}}{\bar{\tilde E}}\left (1-\frac{\bar{\cal
E}}{\bar{\tilde E}}\right )\leq 0$ in the complete accordance to the
general results obtained for the adiabatic case in the previous
section. It is clear, that both total quantum system and its quantum
subsystem considered above have the real physical sense and are
really interesting for all possible applications. Here we would like
to note, that one can generalize the results of this section to the
case of set of the quantum subsystems which form together the closed
total quantum system of the discussing type.

\subsection{Example II: Left-right asymmetry and time arrow}

\setcounter{equation}{0}

In this section we consider a system, which consists of two parts
again (its \, `left' and `right' constituents) -- in accordance to
specification of the symmetry operator $\Gamma$ taken in this case.
Namely, we define it as proportional to some parity operator; this
leads to the double realization $\psi_{\pm n}$ for the all energy
levels $E_n$ . The corresponding eigenvalue problem reads:
\be\label{5} E\,\psi_{\pm}=E_n\,\psi_{\pm}, \qquad
\Gamma\,\psi_{\pm}=\mp\gamma\,\psi_{\pm}; \ee we take the common
basis $\{\psi_{\pm}\}$ of the energy and decay operators in the
orthonormal form (i.e. below we mean that
$\psi^+_{\pm\,n_1}\psi_{\pm\,n_2}= \delta_{n_1n_2}$ and
$\psi^+_{+\,n_1}\psi_{- \,n_2}=0$ for the all indexes $n_1, \,
n_2$). For definiteness, we name the eigenvectors $\psi_{-\, n}$ and
$\psi_{+\,n}$ as the `right' and `left' ones, respectively, and take
$\gamma>0$.

Then, it is clear that the $\tau$-dependent basis vectors
$\Psi_{\pm}=e^{-i{\cal H}_{\pm n}\tau/\hbar}\psi_{\pm}$, where
${\cal H}_{\pm n}=E_n\pm\, i \frac{\hbar}{2}\gamma$, satisfy the
Schr\"{o}dinger's equation identically. Let us consider an arbitrary
solution $\Psi$ of this equation; it can be represented as some
linear combination $\Psi=\sum{_{\pm\, n}}C_{\pm n}\Psi_{\pm n}$ with
the constant coefficients $C_{\pm n}$ in respect to this basis.
Using the projection relation given above, one can calculate the
probabilities $P_{\pm n}$ to find the system which is situated in
the state $\Psi$ in the energy eigenstates $\Psi_{n\,\pm}$. The
result reads: \be\label{6} P_{\pm n}=\frac{w_{\pm n}}{Z}, \ee where
$w_{\pm n}=e^{\pm \gamma t}\,\tilde w_{\pm n}$,\,\, $Z=e^{\gamma
t}\tilde Z_{+}+ e^{-\gamma t}\tilde Z_{-}$;\,\, and also
\be\label{7} \tilde w_{\pm n}=\rho_{\pm n}e^{-E_n\beta}, \qquad
\tilde Z_{\pm}=\sum_n\tilde w_{\pm n}, \ee where $\rho_{\pm
n}=|C_{\pm n}|^2$.

Our goal is to study evolution of the `left' and `right' parts of
the system, which we relate to subsets of the probabilities $P_{-\,
n}$ and $P_{+\, n}$, respectively. It is easy to check, that this
evolution can be expressed in terms of the dynamics of effective
subsystems with the probabilities \be\label{8} \tilde p_{\pm\,
n}=\frac{\tilde w_{\pm \, n}}{\tilde Z}, \ee which have the standard
$\beta$-dependent statistical form in view of Eq. (\ref{7}). Note,
that in the isothermal case (with $\beta=\beta(t)={\rm const}$), one
obtains immediately, that \be\label{9} {\rm lim}_{_{t\rightarrow \pm
\infty}}\,P_{\pm\, n}\left ( t\right )=\tilde p_{\pm\,n}, \qquad
{\rm lim}_{_{t\rightarrow  \pm \infty}}\,P_{\mp\, n}\left ( t \right
)=0.\ee Thus, in the isothermal evolution this quantum system
transforms from its `left' realization to the `right' one. As the
result, \be\label{10} {\rm lim}_{_{t\rightarrow  \pm
\infty}}\bar\Gamma\left ( t\right )=\pm\gamma,\ee i.e. total shift
of the average value of the decay operator $\Gamma$ is equal to
$2\gamma$. It is clear, that the same situation takes place for the
arbitrary thermodynamic regime, which admits the asymptotic inverse
temperatures $\beta_{\pm}={\rm lim}_{_{t\rightarrow
\pm\infty}}\beta(t)$. In this case one must replace $\tilde p_{_{\pm
\, n}}$ by $\tilde p_{_{\pm \, n}}(\beta_{\pm})$ in Eq. (\ref{9}).

Now let us study in details a special case where the quantum system
reaches a total symmetry between its left and right constituents at
some finite time $t_{\star}$. Namely, we are interesting in the
dynamics with $P_{+\,n}( t_{\star})= P_{-\,n} ( t_{\star})$ for all
values of the collective index $n$. Taking $t_{\star}=0$, one
obtains the system with $\rho_{+\,n}=\rho_{-\,n}\equiv \rho_{n}$, so
\be\label{11} P_{\pm\,n}=\frac{\tilde p_n}{1+e^{\mp2\gamma t}}, \ee
where $\tilde p_n=\tilde w_n/\tilde Z$,\, $\tilde Z=\sum _n\tilde
w_n$, and $\tilde w_n=\rho_ne^{-E_n\beta}$. Thus, in this special
case one deals with the single effective system with set of the
$\beta$-dependent probabilities $\tilde p_n$. Then, for the total
system under consideration the average value of the decay operator
is \be\label{12} \bar\Gamma=\gamma\,\tanh\, \gamma t, \ee so the
quantity $\bar\Gamma (t)$ plays a role of the monotonously
increasing Lyapunov function for the attractor `right' state, and
detects the dynamical parity breaking in the system. This function
defines the universal arrow of time, it demonstrates the
irreversible character of the system dynamics in the really
transparent form.

Note, that form of this function does not depend on the concrete
temperature regime $\beta=\beta (t)$. Also, it is interesting to
stress, that the average energy $\bar E$ of the total system
coincides with the same quantity for the effective subsystem
$\bar{\tilde E}=\sum_{_n}\tilde p_n\,E_n$. Thus, in this special
case the isothermal regime $\beta={\rm const}$ and the adiabatic
regime $\bar E={\rm const}$ become identical. To consider another
regime of the evolution, one can take, for example, the symmetry
operator $Q$ with `left' and `right' eigenvalues $Q_{-n}$ and
$Q_{+n}$ which does not coincide identically. Then, the
thermodynamic regime with $\bar Q={\rm const}$ defines the
`temperature curve' with \be\label{13} t=\frac{1}{2\gamma}\,{\rm
ln}\,\left (\frac{\,\,\bar Q\,\,-\bar{\tilde Q}_{-}}{\bar{\tilde
Q}_{+}-\bar Q}\right ), \ee where $\bar{\tilde Q}_{\pm}=\bar{\tilde
Q}_{\pm}(\beta)=\sum_n Q_{\pm \, n}P_{\pm\, n}(\beta)$. For this
regime $\bar{\tilde E}_{,t}=-\beta_{,t}D_{\tilde E}^2\neq 0$, where
$ D_{\tilde E}$ is a dispersion of the energy for the effective
subsystem introduced above.  It is clear, that one can use not only
the symmetry operators for fixing and study of non-isothermal
thermodynamic regimes. Actually, it is possible to relate such
regimes to average values of the observables which does not commute
with the Hamiltonian of the system under consideration.

One important realization of the generalized quantum theory is
related to theory of the massless Dirac field. Namely, it is easy to
see, that the modified Dirac equation \be\label{14} i\hbar
\,\Psi_{,\tau}=\left (\vec\alpha\,\vec p\,-\,i\,
\frac{\hbar}{2}\,\gamma\,\gamma_{5}\right )\Psi \ee has the standard
form with the Dirac energy operator $E=\vec\alpha\,\vec p$ (where
$\vec p$ is the momentum operator), and the parity decay operator
$\Gamma=\gamma\,\gamma_{5}$. Here $\alpha_k=\gamma_0\gamma_k$ \,
$(k=1,2,3)$, and the Hamiltonian commutation relation is satisfied
obviously. Thus, in complete accordance with the general results
presented above, the originally mixed left-right massless Dirac
quantum system transforms to its strictly polarized asymptotic state
in the case of $\gamma\neq 0$. One can choose the sign of this free
constant parameter to obtain appropriate helicity of the final state
to identify it with the real neutrino system.

Actually, it is a well known fact, that the real neutrinos are
`left' (whereas antineutrinos are `right'), and this circumstance
seems intriguing in view of its unclear and, moreover, `pure random'
status in the standard particle physics. We think, that this our new
approach, which provides a dynamical solution for the helicity
asymmetry problem, is more natural than fundamentally asymmetric
standard scheme. Also, the new approach relates the parity violation
in the real neutrino system with the arrow of time in its evolution
(and with the second law of thermodynamics after all). Thus, in
framework of the generalized quantum theory the `random' feature of
the real quantum physical system becomes a consequence of its
irreversible history from the universal thermodynamic point of view.

\subsection{Discussion}

In this part we have developed new thermodynamic generalization of
the quantum theory which possesses a well-defined arrow of time in
the most important regimes of its thermodynamic evolution. This
generalization is based on the use of non-Hermitian Hamiltonian,
which commuting Hermitian and anti-Hermitian parts define the energy
operator and minus half of the operator of decay parameters of the
energy ejgenstates, respectively. Also, this modified quantum theory
deals with the complex time parameter, which real part coincides
with the `usual' time, whereas the imaginary one is identified as
minus half of the inverse absolute temperature. We postulate
strictly analytic dependence of state vectors on the complex time
(in the Schrodinger's picture), and consider the generalized
conservative systems with time- and temperature-independent
Hamiltonian.

This pure Hamiltonian scheme is completed by introducing of the
thermodynamic regime of evolution, which must be defined in the
terms of average value of the corresponding observable quantity. In
fact, this fixation of the thermodynamic regime means
`incorporating' of the macroscopic observer into the originally
microscopic framework of the modified quantum theory.
Constructively, every consistent thermodynamic regime determines
some evolution of temperature of the system, i.e., it defines some
temperature curve on the complex plane of the parameter of
evolution. Our main goal is related to search and study of the
irreversibility effects of the corresponding quantum dynamics. In
this paper we have studied in details the isothermal and adiabatic
regimes of evolution; we have shown that they actually possess the
well-defined arrow of time.

Also, we have studied the problem of thermodynamic evolution of the
quantum subsystem. Namely, we have decomposed the modified quantum
system into two parts - the subspace with extreme value of energy
(i.e., vacuum for the case of minimal energy), and the subsystem
(the quantum subsystem mentioned above) which is constructed from
the all remaining energy eigenstates. We have analyzed the situation
when this subsystem (as a single whole system) and the state with
extreme energy are characterized by different times of the decay. We
have established, that the average energy of the subsystem, as well
as its conventional entropy, are the monotonous functions; they
actually define the arrow of time in the, for example, adiabatic
regime of evolution of the original quantum system.

Also, we have formulated new and concrete mechanism of `insertion'\,
of the second law of thermodynamics into the standard quantum
theory. Namely, it is shown, that to guarantee the functioning of
this law, one needs only in the adding of the positive parameter of
the decay to the vacuum state of the theory. This simple mechanism
gives actually promising general scheme of modernization of the all
quantum physics.

\section{Generalized classical theory}

The main goal of this part is to develop a consistent classical
limit of the quantum theory presented and studied above. This new
classical theory must possess the main property of the original
quantum theory -- the essential irreversibility of its dynamics.
Thus, it must be a classical mechanics (or/and a classical wave
theory) unified with a classical thermodynamics. We derive it, using
asymptotical approach of the Laplace type to calculation of the
classical limit of the average values of the quantum variables and
all possible correlations of them.

At the first time, we establish the consistent classical dynamics of
quantum numbers (like the quantum numbers $n$, $l$ and $m$ in the
quantum solution of the Kepler problem), which possesses all the
fundamental properties of the theory in its general quantum regime
of evolution. In terms of these numbers, as it is shown below, the
Hamiltonian structure of the theory reaches the simplest form.
Namely, it is easy to understood, that if one considers the quantum
numbers as a possible set of the canonical moments, one deals with
the representation which Hamiltonian is coordinate-independent. In
fact, this means study of the problem using some modification of the
Hamilton-Jacoby approach: this representation is based on the use of
the so called `action-angle variables',\, which form the most
convenient set of the canonical coordinates and moments.

In our construction of the classical limit of the quantum theory we
start with the extremal (variational) principle. We define the
classical dynamics as the dynamics of the resonance structures in
the probability density distributions. Namely, we consider the
evolution of well-defined maximums (`resonances') of the probability
density, which were formed at the initial time and cut terms
depending on the Planck constant $\hbar$ (in `hard'\, classical
regime). We would like to stress, that our classical limit of the
modified quantum theory considered in the first part has two
independent infinitesimal parameters: the resonance width
$\varepsilon$ and the Planck constant $\hbar$. In fact, the first
parameter defines the classical kinematical objects (these
`resonances'\, themselves) in the general quantum framework, whereas
vanishing of the second one maps the general quantum dynamics into
its special classical regime. Note, that one of our goals is to save
a control under the dissipation/concentration of the probability
density resonances in the classical theory framework. We are
interesting in building of classical dynamics which allows one to
deal with some remnants of the quantum uncertainty principle and its
various consequences.

From the pure kinematical point of view, we have established, that
the convenient uncertainty relations can be incorporated correctly
into the general structure of the classical theory. These
fundamental relations are expressed in terms of the classical limit
of the correlations between the canonical coordinates and moments.
Also, we have pointed out, that the correlations under discussion
are the analogies of the metric coefficients from the General
Relativity. They describe the curvature effects in dynamics of the
classical objects (`resonances') in the curved phase space. Note,
that it arises a possibility to identify the classical gravitation
as the quantum effect of the dispersion type.

From the pure dynamical point of view, it is shown, that the
classical evolution of the system is really time-irreversible.
Namely, we have proved the presence of the arrow of time in the same
thermodynamical regimes as it had been done in the quantum theory
case. Firstly, we have established this irreversibility in terms of
the action-angle variables, and, after that, we have expanded the
result to the arbitrary canonical variables case. We have detected
the arrow of time not only in the `hard'\, classical limit of the
original quantum theory defined above (with $\hbar=0$). Moreover, we
have proved irreversibility of the theory dynamics in the case of
$\varepsilon\rightarrow 0$ and free constant parameter $\hbar$ (the
`soft'\, classical theory). This 'soft'\, regime seems really
promising in context of the quasi-classical gravity incorporation
into the theory under consideration.

For the `soft'\, variant of the classical theory, it is established,
that the correlations can be unified into the single null-curvature
matrix field of the symplectic type. The corresponding symplectic
structure of the theory remains conserved under the action of the
symplectic group of the continuous canonical transformations. This
group coincides with matrix-valued group of the hidden symmetries of
the standard General Relativity. We have estableshed, that the
generalized Hamilton equations describing our theory in its `soft'\,
regime also possess the hidden symplectic symmetry.

Then, we have studied and classified these canonical symplectic
maps, and extract a non-linear sector of transformations of the
Ehlers type from them. Using one special established discrete map,
we have detected a `duality'\, between the corpuscular and wave
classical sectors in the general `soft'\, classical theory framework
under consideration. Finally, we have proposed, how one can simplify
the constructing and the following study of dynamics of the
resulting classical theory by the help of canonical transformations
found.

\subsection{Modified classical dynamics in action-angle variables}

\setcounter{equation}{0}

The main goal of this section is to show, that in the best
physically motivated quantum theory with $[{\cal H}_1,\,\, {\cal
H}_2]=0$ one has not any problem with derivation of the motion
equations, which describe its classical regime in the consistent
form. In the other words, this means a possibility of construction
of the (modified) classical dynamics, which corresponds the
(generalized) quantum theory studied above. This dynamics must
possess an arrow of time and will have not any artificial
restrictions on the Hamiltonians of the system (like quadratic
structure in respect to the canonical variables, etc.).

For our goals, it is convenient to use the notation $w(t, n),\, p(t,
n)$, etc., for the all $t_{\alpha}$- and $n_{k}$-dependent dynamical
variables. For example, the energy `levels'\, and the inverse times
of decay (the logarithmic decrements of the states) will be written
now as $E(n)$ and $\gamma (n)$ (in view of their time-independent
nature, see the part 1). Thus, in the general quantum regime of the
theory under consideration, one deals with the probabilities \be
p(t, n)=\frac{w(t, n)}{\cal Z}. \ee Here the scale factor is defined
as ${\cal Z}(t)=\sum_n w(t, \, n)$, whereas \be w(t, \,
n)=\rho(n)\exp{\left [-\left (E(n)\beta +\Gamma(n) t\right )\right
]},\ee and the functions $\rho(n)$ are the corresponding weight
parameters. Our main idea is to use the set of quantum numbers $n$
as the new and the most natural set of dynamical variables (the
canonical moments) for this system to detect the classical regime of
its evolution in the most simple and explicitly consistent form.

At the first time, let us stress again, that the collective
parameter $n$ is the $N$-dimensional set of the quantum numbers for
the theory with $N$ degrees of freedom. These quantum numbers are
the integer-valued quantities defined by the corresponding
stationary Schrodinger problem. Of course, we mean some choice of
the complete set of the commuting quantum variables taken (which
include both the Hamiltonian operators), for the correct fixing of
the all indexes $n_k$, $k=1,...,N$. Our approach to constructing of
the classical limit of the quantum theory under consideration
coincides with conventional understanding of the quantum numbers in
the classical limit of the standard quantum theory. Namely, we
define the classical region of the quantum theory as the one with
$n>>1$, where the variable $n$ can be understood as the continuous
quantity with a really high approximation. In particular, we will
differentiate in respect to the constituents of this collective
variable, i.e. in respect to the set of the new canonical variables
$n_k$. In doing so, we are taking into account some well-defined
infinitesimal physical parameters of the theory, which enter into
the corresponding relations, being multiplied to the variations of
the originally discrete variables $n_k$.

Of course, our definition of the classical regime of the quantum
theory is not based on the continualization of the parameter $n$
only. Also, we suppose the presence of the well-defined resonance
structure for the probability density in the representation taken (a
single `probability resonance',\, for the definiteness). This
resonance means, for example, the localization of the object in the
physical space (if one uses the spatial coordinate representation),
and the well-defined wave pocket (for the case of the momentum
representation).

Thus, we are interesting in the study of the classical dynamics of
quantum states with the resonance structure of the probability
density in the space of the collective parameter $n$. Note, that it
is really important to save the corresponding dispersion structure
of the theory: in the rough limit without any dispersions in respect
to the collective parameter $n$, one deals with `pure energetic
solutions'\, of the Schrodinger's equation. It is easy to see, that
they lead to the probabilities $P(n)$ independent of the
evolutionary parameters $t$ and $\beta$. Now let us remember, that
these parameters have the sense of the `usual'\, physical time and
the inverse absolute temperature, respectively, as it was shown in
the part 1. However, such situation is not of any interest in
context of the study of arrow of time problematic. Below we show,
that taking into account of these dispersion parameters in the
classical regime of the quantum theory allows one to save all the
irreversible quantum effects in the framework of new consistent
scheme for the classical dynamics.

Now let us rewrite the weight coefficient $\rho(n)$ in the
exponential form, \be\rho(n)=\exp{\left (-\sigma\left (n\right
)\right )}.\ee Then, for the probability functions one obtains the
following representation: $p(t, n)=w(t, n)/{\cal Z}$, where ${\cal
Z}=\int\, dn\,\, w(t, n)$, and \be w(t, n)=\exp{\left (-{\cal
S}_2\right )}\ee with \be\label{S}{\cal S}_2=
\sigma+\sum_{\alpha}\tilde E_{\alpha}\tilde t_{\alpha}.\ee Here we
have put $\tilde t_1=t,\, \tilde t_2= \beta$, and $\tilde
E_1=\Gamma,\, \tilde E_2=E$ to stress a possibility of
generalization of the corresponding dynamics to the case of the
arbitrary time dimension. Actually, it will be seen, that all the
relations in this dynamics hold their form for any value of the
`time dimension'\, $A$ (where $\alpha=1,... A$). Of course, the
hidden complex nature of the real physical time takes place in the
original case of $A=2$ only.

Then, let us denote a `maximum point'\, of the probability density
for the solution under consideration as $n_c$, i.e.,\be P\left (
t,\,\, n_c\right )= P_{{\rm max}}=\max_{n}P\left ( t,\, n\right
).\ee Here it is important to stress, that below we study only
solutions which possess well-defined `maximum point'\, for the any
time moment $t_{\alpha}$. It is clear, that this `point'\, will be
some function of this collective time parameter $t_{\alpha}$, i.e.,
one deals with $n_c=n_c(\tilde t_\alpha)$. We relate naturally a
`position'\, of the classical object with this extremal value of the
parameter $n$. Thus, we identify the classical theory with the
theory which describes the dynamics of resonance structures of the
probability density in the original quantum theory. In the `most
hard'\, classical limit they are described by the delta-functional
distributions.

It is clear, that the highly-resonance classical character of the
probability density must be guaranteed by corresponding choice of
the weight parameters $\sigma(n)$. Actually, these parameters are
some initial data of the original quantum state, so they can be
chosen in the appropriate way at the beginning of the evolution of
the system. We plan to discuss this choice later. Here the most
important goal is to write down the defining relations for the
`classical position'\, $n_c$ of the object under consideration in
the constructive form. Thus, we must derive the modified Hamiltonian
equations in the quantum theory motivated and consistent form. The
last demanding is not trivial: the equations waited will be partial
differential equations of the first order (we have $A$ independent
dynamical variables in the theory). This means, that the rough cut
of the starting quantum problem can destroy consistency of the
dynamical system under consideration.

We perform the derivation of the modified Hamiltonian equations in
the $n$-variables by the help of differentiation of the conventional
extremum relation in respect to the total set of times $\tilde
t_{\alpha}$: \be\label{diff} \frac{d}{d\tilde t_{\alpha}}\left
(\frac{\partial {\cal S}_2\left (\tilde t_{\alpha},\, q\right
)}{\partial q_{m}}\right )_c=0,\ee where the index $c$ means the
substitution $q=q_c$ into the differentiation result. Note, that the
$\tilde t_{\alpha}$-differentiation is understood here in respect to
the total $\tilde t_{\alpha}$-dependence of the corresponding
quantity (i.e. with taking into account the hidden dependence
$n_c=n_c(\tilde t_{\alpha})$ in this equation). It fact, Eq.
(\ref{diff}) means a conservation of the extremal character of the
classical trajectory of the system during its possible physical
evolution. Actually, the quantity in the brackets in (\ref{diff}) is
equal to zero for the `maximum point' of the probability density.

Then, it is not difficult to prove, that the dynamical equation for
the $n=n_c$ quantity, which follows from Eqs. (\ref{S}) and
(\ref{diff}), reads: \be\label{n} n_{,\,\tilde t_{\alpha}}=-{\cal
A}_2^{-1}\dot {\tilde E}_{\alpha},\ee where \be{\cal A}_2=\ddot{\cal
S}_{2\,c}=\ddot{\sigma}_{c}+\sum_{\beta}\ddot{\tilde
E}_{\beta\,\,c}\tilde t_{\beta }\ee Here $\dot {\tilde E}_{\alpha}$
is the N-column with the elements $\partial \tilde
E_{\alpha}/\partial n_{k}$, whereas $\ddot{\sigma}_c$ and
$\ddot{\tilde E}_{\alpha\,c}$ are the $N\times N$ symmetric matrices
with the coefficients $\partial^2 \sigma/\partial n_{k}\partial
n_{l}$ and $\partial^2 \tilde E/\partial n_{k}\partial n_{l}$
(calculated at the `classical trajectory'\, $n=n_c$), respectively.
Our main statement is that this dynamical system of partial
differential equations is consistent for the arbitrary functions
$\sigma (n)$ and $\tilde E_{\alpha}(n)$ taken as initial data of the
solution. This means, that in the $n$-representation our dynamical
problem has not any additional and artificial restrictions -- in
complete agreement with the announcement given at the beginning of
this part of the article. The only restriction which must be put on
the system dynamics is the original commutation relation for the
Hamiltonians of the theory, in view of the starting general quantum
solution explored. Note, that we mean $n$ as the column of the
height $N$ here and in the all following relations.

It is really important to stress, that the functions $E_{\alpha}(n),
\, \sigma(n)$ do not depend on the Planck constant $\hbar$. Thus, in
the classical regime our quantum relations have not any `rule
parameter',\, related to this actually small physical quantity.
Furthermore, we need in a new `rule parameter'\, to define a correct
classical theory derivation. It is clear, that it can be found for
the macroscopical system as some proportionality coefficient, which
defines a classical scale for its `big'\, mass, charges, etc. In the
other words, there is not any reason for extracting of the quantum
theory with its classical regime using only the limit procedure
$\hbar\rightarrow 0$. We relate this resonance structure of the
kinematics of the theory to the natural parameter $\varepsilon$ such
that the classical limit for the functions $\bar E, \, \bar \sigma$,
etc., must be proportional to the quantity $1/\varepsilon$.

In the next section, it will be shown that the limit procedure
$\varepsilon\rightarrow 0$, together with the procedure, defined by
the relation $\hbar\rightarrow 0$, is enough for the correct
formulation of the classical theory. Here it is necessary to note,
that the limit at $\hbar\rightarrow 0$ is important to calculation
of the set of canonical moments (their average values)
$p_k=-i\hbar\partial/\partial_{n_k}$ in the theory. The result
reads: \be p_k=\left (\frac{\partial {\cal S}_1\left (\tilde
t_{\alpha},\, q\right )}{\partial q_{m}}\right )_c,\ee where we have
combined $N$ moment variables to the single N-column quantity. In
this formula ${\cal S}_1$ means the real part of the phase ${\cal
S}$ of the wave-function $\Psi$ in the $n$-representation of the
theory, which is defined by the relation \be \Psi (n,
t)=C(n)\exp{\left ( \frac{i}{\hbar}\,{\cal S}\right )}.\ee It is
easy to calculate its explicit form in the situation under
consideration; the result reads: \be {\cal S}_1(t,\,
n)=\lambda(n)+E(n)t-\frac{\hbar^2}{4}\Gamma(n)\beta.\ee Here we mean
again the original theory with the double time parameter, and define
the function $\lambda (n)$ according to the relation \be
C(n)=\sqrt{\rho (n)}\exp{\left (-\frac{i}{\hbar}\lambda \right
)}.\ee

Then, in the explicit form, \be p=\dot{\lambda}+\dot{E} t_1,\ee and
for the time derivatives of the moment column $p$ on obtains the
equation $p_{\tilde t_{\alpha}}=\dot {E}\delta_{1,\alpha}+{\cal
A}_1n_{{,\tilde t}_{\alpha}}$, where \be{\cal A}_1=\ddot{\cal
S}_{1\, c}=\ddot{\lambda}+\ddot{E}\,t.\ee Thus, using the relation
(\ref{n}), one concludes, that \be\label{p} p_{,\tilde
t_{\alpha}}=\dot {E}\delta_{1,\alpha}+{\cal A}_1{\cal
A}_2^{-1}\dot{E}_{\alpha}.\ee It is clear, that the relations
(\ref{n}) and (\ref{p}) form the pair of Hamiltonian equations for
the theory under consideration. The main statement is that this
system of equations is consistent, i.e., that the mixed time
derivatives for coordinate and moment variables are equal to the
inverse one. This means, that this system of classical equations is
a correct one.

Our second statement is related to irreversibility of evolution of
the system, defined by the Hamiltonian Eqs. (\ref{n})--(\ref{p}),
combined with some specified temperature regime $\beta=\beta(t)$. We
consider below isothermal and adiabatic regimes again, as it had
been performed in the quantum part of this work. To prove the
presence of arrow of time in these physically important regimes, let
us choose a representation, where $\Gamma=\Gamma(n_1)$, and
$E=E(n_2)$. It is clear, that this representation exists for the
theory taken (i.e., for the theory with commuting Hamiltonian
operators). From the Hamiltonian equations it follows, that
\be\label{syst} \ba {ccc} n_{1, t}=-\Gamma^{'}({\cal
A}_2^{-1})_{11}, &\quad& n_{2, t}=-\Gamma^{'}({\cal
A}_2^{-1})_{21},\\n_{1, \beta}=-E^{'}({\cal
A}_2^{-1})_{12},&\quad&n_{2, \beta}=-E^{'}({\cal
A}_2^{-1})_{22},\ea\ee where the prime means the derivative in
respect to corresponding single variable of the function under
consideration. We take the function $\Gamma$ as the waiting Lyapunov
one, according to the analysis performed in the quantum part of this
work. Then, for the total $t$-derivative of this function, i.e. for
the quantity
$\Gamma_{;t}=\Gamma^{'}n_{1;t}=n_{1,t}+\beta^{'}n_{1,\beta}$, one
can calculate its explicit form using Eqs. (\ref{syst}). For the
isothermal case ($\beta={\rm const}$) the result reads: \be
\Gamma_{;t}=-\left (\Gamma^{'}\right )^2\left ({\cal A}_2^{-1}\right
)_{11}.\ee It is clear, that $\Gamma_{;t}<0$. In the adiabatic
situation with $E={\rm const}$ one obtains \be
\beta^{'}=-\frac{\Gamma^{'}}{E^{'}}\frac{\left ({\cal
A}_2^{-1}\right )_{21}}{\left ({\cal A}_2^{-1}\right )_{22}}\ee for
its temperature regime, and \be\Gamma_{;t}=-\frac{\left
(\Gamma\right )^2}{\left ({\cal A}_2^{-1}\right )_{22}}\left
|\ba{ccc}\left ({\cal A}_2^{-1}\right )_{11}&\,&\left ({\cal
A}_2^{-1}\right )_{12}\\\left ({\cal A}_2^{-1}\right )_{21}&\,&\left
({\cal A}_2^{-1}\right )_{22}\ea\right |\ee for the total time
dependence of the Hamiltonian function $\Gamma$. It is clear, that
in the both thermodynamical regimes considered here one deals with
decreasing evolution of the function $\Gamma(t)$ for the arbitrary
initial data taken. This proves the Lyapunov nature of this quantity
and the presence of arrow of time in the dynamics of the classical
system under investigation. Note, that in the both situations
considered the inequality $\Gamma_{;t}<0$ is guaranteed by the
positive definition of the matrix ${\cal A}_2$, which we have
supposed for the solution of the dynamical equations to deal with
maximum of the quantity ${\cal S}_2$ on the whole `classical
trajectory'\, $q=q_c(t)$ of the system.

It is necessary to stress, that one has a possibility of taking of
any value of the matrix field $\ddot{\sigma}(n)$, which provides the
resonance-like character of initial data and the resonance-dominated
dynamical history of the probability density $P(t,\, n)$ for the
classical object. In fact, this symmetric matrix field modifies the
dynamics of this object in the same way as a curved metrics `acts'\,
on the trajectory of the point mass (or on the bosonic string, in
view of the two-dimensional nature of the problem under
consideration). The only difference is related to nature of the
dynamical variables in these two theories. Actually, here we deal
with the `action-angle variables'\, of the system, whereas in the
standard mechanics and string theory the corresponding quantities
are the usual physical coordinates. Also, both our parameters of the
dynamics are time-like ones (in the case of $A=2$), whereas their
string theory analogies have the time-like and space-like nature.
From formal point of view, our classical dynamics is the most
closely related to the dynamics of the bosonic string in the
background metric field. In the our case, this `gravitation'\, is
modeled by the matrix $\ddot{\sigma (n)}$, which describes the
`curvature effects'\, in the N-dimensional space of the canonical
variables $n_k$.

At the end of this section let us note, that one cannot take the
initial value $n_{0}=n_{c\, 0}$ of the `classical object position'\,
and the `phase space metric'\, $\ddot{\sigma (n)}$ in the
independent way. Actually, $n_0$ denotes the `maximum pojnt'\, of
the probability density distribution at the `initial times'\,
$\tilde t_{\alpha}$, whereas the `metric'\, $\ddot{\sigma (n)}$
defines this `metric'\, completely. Thus, it is obvious an existence
of the relation between $n_0$ and $\ddot{\sigma (n)}$. This means,
that in the modified Hamiltonian dynamics under consideration
initial data must be consistent with the dynamical characteristics
of the system.

\subsection{Modified Hamilton equations}

\setcounter{equation}{0}

Below we use an experience obtained in the study of the modified
classical dynamics using the action-angle variables, for
investigation of the theory in terms of its original canonical
coordinates $q$ and moments $p$. Our first goal is to derive the
corresponding equations of motion of the Hamilton type for this
theory and to check their consistence. Realization of this program
seems really important in view of the highly non-trivial procedure
of search of the action-angle variables $n$ for the system in the
general case.

Starting this realization, we define the (hard) classical regime for
the quantum theory as the double limit at $\varepsilon\rightarrow 0$
and $\hbar\rightarrow 0$, which is taken in the appropriate way in
the all exact quantum relations. The main facts, which we have
established in the study of classical dynamics of the system in
terms of the action-angle variables, are collected in the
decompositions \be\label{ham1}{\cal S}={\cal
S}_1+i\frac{\hbar}{2}{\cal S}_2, \quad {\cal H}={\tilde
E}_2-i\frac{\hbar}{2}{\tilde E}_1, \quad {\tau}={\tilde
t}_1-i\frac{\hbar}{2}{\tilde t}_2,\ee where all the quantities
written are the classical ones (i.e., they are $\hbar$-free ones; in
other words -- they are the zero $\hbar$-power magnitudes). We again
use the notation $\tilde E_1=\gamma, \, \tilde E_2= E, \, \tilde
t_1=t, \, \tilde t_2=\beta$, and parameterize the wave-function
$\Psi$ in terms of the complex phase ${\cal S}$ as
$\Psi=\exp{(i{\cal S}/\hbar)}$. Thus, we plan to derive dynamical
equations of the classical regime of the quantum theory under
consideration, using the $\hbar$-power structure of the theory
constituents established in the study of its $n$-representation.

Note, that one can restrict this analysis by the study of the single
limit $\varepsilon\rightarrow 0$ only. In the corresponding
quasi-classical theory, one deals with well-localized objects, but
the dynamics of theory contains the free parameter $\hbar$. It is
really interesting to note, that in this variant of the theory one
obtains a dynamical model with highly symmetric algebraic structure
of the symplectic type. We develop in details corresponding general
formalism in the next sections of this work.

The Schrodinger's equation $i\hbar\Psi_{,\,\tau}={\cal H}\Psi$ and
the analytical condition $\Psi_{,\,\tau^*}=0$ leads one to the
following relations: \be\label{ham2}{\cal S}_{1,\,\tilde
t_1}+i\frac{\hbar}{2}{\cal S}_{2,\,\tilde t_1}&=&-\left (\hat{\tilde
E}_2-i\frac{\hbar}{2}\hat{\tilde E}_1\right ),\nonumber\\{\cal
S}_{1,\,\tilde t_2}+i\frac{\hbar}{2}{\cal S}_{2,\,\tilde
t_2}&=&i\frac{\hbar}{2}\left (\hat{\tilde
E}_2-i\frac{\hbar}{2}\hat{\tilde E}_1\right ).\ee Here, the hatted
quantities are defined as \be\label{ham3}\hat{\tilde
E}_{\alpha}=\Psi^{-1}\tilde E_{\alpha}\Psi,\ee and mean nontrivial
functions of the coordinates $q$ and some set of the derivatives of
the complex phase ${\cal S}$. The last dependence originates from
the $p=-i\hbar
\partial/\partial q$-terms in the operators $\tilde E_{\alpha}$.

To `feel'\, a nature of the hatted quantities, it is convenient to
calculate the simplest polynomial moment functions. In fact, we need
in the linear and quadratic ones only for the consistent development
of the `hard'\, classical limit of the theory. Performing
straightforward calculations, it is not difficult to prove, that
\be\label{ham4}\hat p_k={\cal S}_{,q_k},\qquad \hat{p_kp_l}={\cal
S}_{,q_k}{\cal S}_{,q_l}-i\hbar{\cal S}_{,q_kq_l}.\ee Using these
relations, one can prove, that \be\label{ham5} \hat{\tilde
E}_{\alpha}=\tilde E_{\alpha}(q,\,{\cal
S}_{1,\,q})+i\frac{\hbar}{2}\left [\tilde E_{\alpha,\, p^T}{\cal
S}_{2,\, q}-{\rm tr}\left (\tilde E_{\alpha,\, pp^T}{\cal S}_{1,\,
qq^T}\right )\right ],\ee up to $\sim\hbar$ terms. Note, that we
explore the compact collective notation, which can be `decoded'\, as
$\tilde E_{\alpha,\, p^T}{\cal S}_{2,\, q}=\sum_k{\tilde
E_{\alpha,\, p_k}{\cal S}_{2,\, q_k}}$ and ${\rm tr}\left (\tilde
E_{\alpha,\, pp^T}{\cal S}_{1,\, qq^T}\right )=\sum_{k\,l}\tilde
E_{\alpha,\, p_kp_l}{\cal S}_{1,\, q_kq_l}$, etc., in the
corresponding relations.

Then, from Eqs. (\ref{ham2}) it follows, that the relations, which
define the action-like quantities ${\cal S}_{\alpha}$ in the regime
taken, read: \be\label{ham6}{\cal S}_{1,\, \tilde t_1}&=&-\tilde
E_{2}(q,\,{\cal S}_{1,\,q}),\nonumber\\ {\cal S}_{1,\, \tilde
t_2}&=&0,\ee and \be\label{ham7}{\cal S}_{2,\, \tilde t_1}&=&\tilde
E_{1}(q,\,{\cal S}_{1,\,q})-\tilde E_{2,\, p^T}(q,\,{\cal
S}_{1,\,q}){\cal S}_{2,\, q}+{\rm tr}\left [\tilde E_{2,\,
pp^T}(q,\,{\cal
S}_{1,\,q})\,\,{\cal S}_{1,\, qq^T}\right ], \nonumber \\
{\cal S}_{2,\, \tilde t_2}&=&\tilde E_{2}(q,\,{\cal S}_{1,\,q}).\ee
It is seen, that the demanding of $\hbar\rightarrow 0$ leads to the
absolutely non-symmetric structure of the dynamical equations. In
these equations, roles of the operators $E$ and $\Gamma$ are
non-identic transparently. This fact follows from the absolutely
asymmetric $\hbar$-dependence of the right sides of the relations
(\ref{ham2}).

Let us discuss the general physical sense of the system
(\ref{ham6})-(\ref{ham7}). First of all, from the second equation of
(\ref{ham6}) it follows that ${\cal S}_1$ does not depend on the
inverse temperature $\beta$, i.e., ${\cal S}_1={\cal S}_1(t,\, q)$.
Then, the first equation (\ref{ham6}) is the standard
Hamilton-Jacoby equation exactly (because $\tilde E_2$ is the energy
dynamical variable -- the standard Hamiltonian of the system). This
means, that ${\cal S}_1$ is a classical action ${\cal S}_a={\cal
S}_a(t,\, q)$ of the system, which satisfies the Hamilton-Jacoby
equation \be\label{ham6'}{\cal S}_{a,\, t}=-E\left (q,\,{\cal
S}_{a,\,q}\right).\ee Then, the physical sense of the function
${\cal S}_2$ can be obtained using the study of the system
(\ref{ham7}). Actually, from the second equation of (\ref{ham7}) it
follows, that \be\label{entr1}{\cal S}_2={\cal S}_e+E\beta,\ee where
the function ${\cal S}_e={\cal S}_e(t,\, q)$ does not depend on
$\beta$. This new `action-like function'\, satisfy the equation
\be\label{entr2}{\cal S}_{e,\, t}=\gamma(q,\,{\cal S}_{a ,\,q})+{\rm
tr}\left [\tilde E_{2,\, pp^T}(q,\,{\cal S}_{a,\,q})\,\,{\cal
S}_{a,\, qq^T}\right ],\ee as it can be easily verified, using the
first equation of (\ref{ham7}) and the definition of ${\cal S}_e$
given by Eq. (\ref{entr1}).

Our statement is: the quantity ${\cal S}_e$, being calculated on the
`classical trajectory'\, $q=q_c$, coincides with the classical
entropy of the system. This means, that the function
$S_e(t,\,q_c(t,\beta))$ is the standard classical entropy, which
corresponds to the system dynamics in the conventional thermodynamic
framework. To prove this `shocking statement',\, one must take into
account, that the classical entropy can be calculated using the
standard relation \be\label{entr3}{\cal S}_e=-{\rm ln}\,{\cal
Z}-\beta\,E,\ee and that in the classical limit
$\varepsilon\rightarrow 0$ one obtains \be\label{entr4}{\cal
Z}=\exp{\left (-{\cal S}_{2\,c}\right )}.\ee Finally, one concludes,
that the modified `hard'\, classical dynamics is completely
described in terms if the classical action and entropy function of
the theory. The incorporation of the entropy into the general
Hamiltonian formalism allows one to predict an arising of the arrow
of time in the corresponding dynamics.

Namely, the main statement, related to the system
(\ref{ham6})-(\ref{ham7}), is that this system is actually
consistent. This property is not obvious, because the system was
derived by the help of the perturbation expansion in respect to the
Plank constant. Then, `cut'\, of the all $\hbar$-dependent terms
performed during a derivation of the system
(\ref{ham6})-(\ref{ham7}) could destroy the consistency of the
original quantum equations. Thus, one must compare an equivalence of
the mixing time derivatives to be sure in correctness of the all
dynamical results established during the study of this theory.

Proof of the consistence under discussion can be performed using the
corresponding  straightforward calculations. First of all, the
identity \be\label{ham8}{\cal S}_{1,\, \tilde t_1\tilde t_2}={\cal
S}_{1,\, \tilde t_2\tilde t_1}\ee is evident in view of the form of
the second equation from the subsystem (\ref{ham6}) (it is easy to
see, that both these mixed derivatives are equal to zero). Then, in
the case of the equations (\ref{ham7}) one has not a subsystem,
because the defining ${\cal S}_2$ relations include the quantities
related to ${\cal S}_1$ in their right sides. However, using the
total system (\ref{ham6})-(\ref{ham7}), one concludes, that
\be\label{ham9}{\cal S}_{2,\, \tilde t_1\tilde t_2}={\cal S}_{2,\,
\tilde t_2\tilde t_1}=-\tilde E_{2,\, p^T}\tilde E_{2;\, q}.\ee This
means, that both the action-like quantities ${\cal S}_{\alpha}$ are
defined by the pure {\it classical}\, system of the equations
correctly, i.e. in the consistent form. This means, that the
classical theory under construction {\it exists}, because it is
defined completely by the equations (\ref{ham6})-(\ref{ham7}).
Actually, we relate this classical theory with some special
solutions of this consistent system of partial differential
equations -- with such solutions, which allow the highly resonance
character in terms of the function ${\cal S}_2$. Taking the
resonance structure of this type as initial data, one can study its
consistent dynamics using corresponding conclusions of Eqs.
(\ref{ham6})-(\ref{ham7}). These conclusions must have sense of the
modified Hamilton equations. Now we start their derivation.

This derivation is based on the use of the same approach, as the one
used for the study of classical dynamics of the system under
consideration in terms of its action-angle variables. Namely, we
start with differentiation of the extremum relation
\be\label{ham10}\left ({\cal S}_{2,\,q_k}\right )_c=0\ee in respect
to the time variables $\tilde t_{\alpha}$. Here $({\cal
S}_{2,\,q_k})_c={\cal S}_{2,\,q_k} (t,\, q(t))$, where $q(t)$ means
the classical `trajectory', \, which minimizes the ${\cal S}_2(t,
q)$ value. Starting from this point, let us come back to the
physical notation $t, \beta$ and $E, \Gamma$ in the all relations.
Performing the calculations, one concludes, that \be\label{ham11}
q_{,\,t}&=& E_{,p}-{\cal
A}_2^{-1}\Gamma_{*;q},\nonumber\\q_{,\,\beta}\!&=&\! -{\cal
A}_2^{-1}E_{;q}, \ee where the index `$;q$'\, is used to denote the
total $q$-derivative taken, and we have introduced the
`renormalized'\, decay function $\Gamma_{*}$ as:
\be\label{ham12}\Gamma_{*}=\Gamma+{\rm tr}\left [\tilde E_{2,\,
pp^T}(q,\,{\cal S}_{1,\,q})\,\,{\cal S}_{1,\, qq^T}\right ].\ee
Then, for the moments $p={\cal S}_{1,\,q}$, the corresponding
defining relations read: \be\label{ham13} p_{,\,t}&=& -E_{,\,
q}-{\cal A}_1{\cal
A}_2^{-1}\Gamma_{*;q},\nonumber\\p_{,\,\beta}\!&=&\! -{\cal
A}_1{\cal A}_2^{-1}E_{;q}.\ee

Note, that the $t$-dependence of the canonical coordinates and
moments includes the traditional `pure Hamiltonian'\, part. However,
the complete right sides of the motion equations are modified by the
presence of dispersion-like structures hidden in the quantities
${\cal A}_{\alpha}$. Eqs. (\ref{ham11}), (\ref{ham12}) describe,
together with the system (\ref{ham6})--(\ref{ham7}), all the
`hard'\, classical dynamics of the system under consideration. This
dynamics can be named as `modified Hamiltonian dynamics';\, our main
goal is to study its irreversibility properties.

However, the initial point in the study of this system of partial
differential equations must concern its consistence. First of all,
it is not difficult to prove, that the moment relations
(\ref{ham13}) are consistent if the ones (\ref{ham11}) for the
canonical coordinates allow this property. To establish this fact,
it is convenient to take into account, that these systems are
related as \be\label{ham14} p_{,\,t}= -E_{;\, q}+{\cal
A}_1q_{,\,t},\quad p_{,\,\beta}= -{\cal A}_1q_{,\,\beta}.\ee The
proof of the equivalence $q_{,\,t\beta}=q_{,\,\beta t}$ is more
difficult. Nevertheless, this consistence condition is actually take
place. Using straightforward calculations, it is possible to show,
that \be\label{ham15} q_{,\,t\beta}-q_{,\beta t}=\left ({\cal
S}_{2,\,t\beta}-{\cal S}_{2,\,\beta t}\right )_{;\,q},\ee so the
consistency of the Hamiltonian equations follows from the
consistency for the phase quantities ${\cal S}_{\alpha}$. Thus, our
double limit $\varepsilon\rightarrow 0, \, \hbar\rightarrow 0$ taken
in the consistent original quantum dynamics leads to the correct
classical regime of the system under consideration. We name it as
`modified Hamiltonian dynamics'\, in view of the obvious reasons.

Now let us consider the problem of presence of the arrow of time in
the modified Hamiltonian dynamics formulated above. First of all,
the total $t$- and $\beta$- derivatives, calculated for the energy
and decay operators $E$ and $\Gamma$, read: \be\label{ham16} \ba
{ccc} E_{,\,t}=-\left ( E,\, E\right ), &\,&  E_{,\,\beta}=-\left (
E,\, E\right ),\\ \Gamma_{,\,t}=-\left ( \Gamma,\, \Gamma\right
)+\left \{ E,\,\Gamma\right \}, &\,& \Gamma_{,\,\beta}=-\left (
\Gamma,\, E \right ). \ea \ee Here, for example, the scalar product
of the quantities $\Gamma$ and $E$ is denoted as $\left ( \Gamma,\,
E \right )$, and defined according to the relation \be\label{ham17}
\left ( \Gamma,\, E \right )=\Gamma_{;\, q^T}{\cal A}_{2}^{-1}E_{;\,
q}. \ee Then, the term $\{ E,\,\Gamma \}$ means the standard
classical Poisson brackets. Below we suppose, that (as in the
quantum theory case) the commutation  relation \be\label{ham18}\{
E,\,\Gamma \}=0\ee takes place for the theory under consideration.
Note, that one must take into account the $total$ dependence on $q$
in the calculation of the derivative $;\,q$ of the corresponding
dynamical quantity. Also, it is important to remember, that the
matrix ${\cal A}_2$ is positively defined, so the scalar production
of the any (nontrivial) dynamical quantity to itself is the positive
magnitude always.

For the total $t$-dependence, i.e., in the case of the temperature
curve $\beta=\beta(t)$ defined, one deals with the decay dynamical
variable, which satisfies the following relation:
\be\label{ham19}\Gamma_{;\,t}=\Gamma_{,\,t}+\beta^{'}\Gamma_{,\,\beta}.\ee
For the isothermal regime of evolution of the system (where
$\beta(t)={\rm const}$) one obtains, that
\be\label{ham20}\Gamma_{;\,t}=-\left ( \Gamma,\, \Gamma\right )<0.
\qquad {(\rm isothermal \,\, case)}\ee Thus, in the isothermal
regime the decay variable $\Gamma$ decreases for any initial data
taken. This means the presence of the arrow of time in the
corresponding dynamics of the system. Then, in the adiabatic case
(where $E_{;\,t}=E_{,\,t}+\beta^{'}E_{,\,\beta}\equiv 0$), one
obtains, that \be\label{ham21}\beta^{'}=-\frac{\left ( E,\,
\Gamma\right )}{\left ( E,\, E\right )}.\ee In derivation of this
and previous results we have used Eq. (\ref{ham16}). After the
substitution of Eq. (\ref{ham21}) into Eq. (\ref{ham19}) and
additional use of Eq. (\ref{ham16}), we finally obtain, that
\be\label{ham22}\Gamma_{;\, t}=-\frac{\left ( \Gamma,\, \Gamma\right
)}{\left ( E,\, E\right )}\left ( 1- \frac{\left ( E,\, \Gamma\right
)^2}{\left ( E,\, E\right )\left ( \Gamma,\, \Gamma\right )}\right
).\ee Note, that in the adiabatic case one has again the evolution
with \be\label{ham23}\Gamma_{;\, t}<0,\qquad {(\rm adiabatic \,\,
case)}\ee as it follows from Eq. (\ref{ham22}) and from the
inequality of Cauchy-Bounyakowsky. Thus, in the adiabatic case we
deal also with the time-irreversible classical dynamics.

Note, that all the results of this section exactly correspond to the
ones obtained in the quantum part of this article. Namely, the only
concretization is related to explicit realization of the scalar
products used in the calculations. Also we would like to stress,
that Eq. (\ref{ham12}) defines the classical value of the decay
variable $\Gamma$ correctly. This follows from the `expected form'\,
of the relations (\ref{ham16})--(\ref{ham23}), presented above,
which are written down in terms of this quantity. Thus, in the
classical limit the decay dynamical variable obtains the `quantum
shift'\, expressed in the form of trace term presented above. This
term does not destroy the irreversible structure of the theory under
consideration.

\subsection{Uncertainty relations in classical theory}

\setcounter{equation}{0}

In this section, we clarify a sense of the quantities ${\cal
A}_{\alpha}$ which had arisen during the study of the classical
dynamics of the system. Our goal is to show, that these variables
are closely related to dispersion characteristics of the theory. The
main statement of this section is that the `soft'\, classical regime
possesses famous uncertainty relations which are well known in the
quantum theory framework. It is shown, that this remarkable fact
takes place on the kinematic level of the classical theory. The
`soft character'\, of the statement is really natural, because in
the `hard'\, regime with $\hbar=0$ one obtains the trivial classical
identity from the original quantum inequalities. However, the simple
dispersion sense of the quantities ${\cal A}_{\alpha}$ conserves
also in the `hard'\, variant of the classical theory too.

First of all we would like to stress, that the average value $\bar
f$ of the quantum variable $f=f(t_{\alpha},\, q)$ we calculate using
the standard relation \be\label{aver}\bar f=\int dq \rho f,\ee where
$\rho=w/{\cal Z}$. In the general operator case with
$f=f(t_{\alpha},\, q, \, p)$, where $p$ is the column of the height
$N$, constructed from the standard canonical moment operators, we
introduce the convenient quantity $\hat f=\Psi^{-1} f \Psi$. Then,
for the average value $\bar f$ of this general quantum variable, one
can use the relation (\ref{aver}), modified by the substitution $f
\rightarrow \hat f$ into its integral term.

For the case of `soft'\, classical regime we suppose only, that the
theory describes some point-like (or corpuscular) object. This
means, that the probability density has the `resonance-like'\,
maximum on the classical trajectory of the system. Our goal is to
calculate all the fundamental dispersion characteristics (i.e., all
the correlations for the total set of the canonical variables) for
this system. It is clear, that these correlations describe a
behavior of the point-like object under consideration at the nearest
vicinity of its classical trajectory in the phase space.

We extract from the solution taken the small parameter
$\varepsilon$, which defines the resonance character of the
probability density of this solution. Then, in the limit of
$\varepsilon\rightarrow 0$ we lead to the `soft'\, variant of the
classical theory under discussion. Performing this limit procedure,
for the scale factor ${\cal Z}$, one obtains the following result:
\be\label{Z} {\cal Z}=e^{-2S_{2\,c}}\sqrt {\frac{\left (\pi\right
)^N}{{\rm det}\,{\cal A}_2}}\left (1+o\left ( \varepsilon\right
)\right ),\ee where the well-known Laplace integral theorem had been
applied, and the notations defined in the previous section had been
used. Below we neglect all the terms $o(\varepsilon)$ in respect to
the leading asymptotic in the all relations. However, for
calculation of correlations of the canonical coordinates and
moments, it is important to save in the asymptotic decompositions
all the terms up to $\varepsilon$. In view of this reason, the main
work relation for the average quantity $\bar f$ reads:
\be\label{middle} \bar f=\hat f_c+\frac{1}{4}{\rm tr}\left
(\ddot{\hat f}_c{\cal A}_2^{-1}\right ).\ee Here $\ddot{\hat f}$ is
the matrix with the coefficients $\ddot{\hat f}_{c\,mn}=(\hat
f_{;\,q_mq_n})_c$. Note, that the index `$_{;\,q_m}$'\, means again
the total derivative of the corresponding dynamical variable in
respect to the canonical coordinate $q_m$.

It is important to stress, that in Eq. (\ref{middle}), one takes
into account all terms up to quadratic ones for the both functions
${\cal S}_2$ and $\hat{f}$, decomposed to the Tailor series in the
surrounding of the `extremal point'\, $q=q_c$. Actually, all the
higher derivative terms form $o(\varepsilon)$ modes; we neglect them
as it was written above. In fact, here we restrict our study by
consideration of the multidimensional `Gaussian bell'\, around the
`classical trajectory'\, of the system. Namely, we perform the
representation \be\label{bell1}{\cal S}_2={\cal S}_{2\,
c}+\frac{1}{2}\left (q-q_c\right )^T{\cal A}_2\left (q-q_c\right
)+\dots,\ee where it was used the extremal condition $(\dot{\cal
S}_{2})_c=0$. Thus, the exponential kernel in the integral form,
which defines the average values, reads: \be\label{bell2}e^{-{\cal
S}_2}\cong \exp{\left [-S_{2\,c}-\frac{1}{2}\left (q-q_c\right
)^T{\cal A}_2\left (q-q_c\right )\right ]},\ee where ${\cal A}_2$ is
the positively defined matrix. This last relation, being
approximated by its leading term at the right side, defines the
$N$-dimensional Gaussian bell. In this connection, the correlation
parameters of the system are closely related to the `bell's
parameters'. In particular, we suppose, that the matrix ${\cal A}_2$
and the `resonance parameter'\, $\varepsilon$ are related as ${\cal
A}_2\sim 1/\varepsilon$, so in the limit case of
$\varepsilon\rightarrow 0$, $\varepsilon\neq 0$, one obtains the
asymptotic results (\ref{Z}) and (\ref{middle}). Note, that our
`hard'\, variant of the modified classical dynamics based on the use
of the terms up to the terms $\sim \varepsilon$, see Eq.
(\ref{middle}).

To develop the general formalism, which describes the classical
limit of the dispersion characteristics of the theory, let us define
the correlation $f\circ g$ of the quantum variables $f$ and $g$
according to the relation \be f\circ
g=\overline{fg+gf}-2\overline{f}\overline{g},\ee where one must
calculate the right side up to the $\sim \varepsilon$ accuracy.
These quantities, being introduced to the theory, describe
non-equivalence of the average value of the product of the dynamical
variables and the product of the average values of them. It is easy
to see, that the squared dispersion ${\cal D}_f^2$ of the dynamical
variable $f$ is related to the correlation quantity $f\circ f$ as
\be {\cal D}_f^2=\frac{1}{2}f\circ f. \ee Thus, a calculation of the
correlations is equivalent to the calculation of the dispersion-like
characteristics of the system. Our goal is to derive the classical
limit of the quantum correlations for the canonical coordinates and
moments.

To perform this work, we use the Laplace integral decomposition
formula (\ref{middle}). The result read: \be\label{correl} \left
(q_m\circ
q_n\right )_c&=&\left ({\cal A}_2^{-1}\right )_{mn}, \nonumber \\
\left (q_m\circ p_n\right )_c&=&\hbar\left ({\cal A}_2^{-1}{\cal
A}_1\right )_{mn}, \nonumber \\ \left (p_m\circ p_n\right
)_c&=&\hbar^2\left ({\cal A}_2+{\cal A}_1{\cal A}_2^{-1}{\cal
A}_1\right )_{mn}. \ee Thus, we have proved the statement formulated
at the beginning of this section: the quantities ${\cal A}_{\alpha}$
actually define all fundamental correlations in the classical theory
under consideration. These quantities, as it is seen from Eqs.
(\ref{ham11}), (\ref{ham13}), enter to the right sides of the
Hamiltonian equations. Thus, the `hard'\, variant of classical
dynamics under consideration takes into account classical limit of
the dispersion characteristics of the system.

It is important to note, that one can use an alternative momentum
representation of the original quantum theory. This representation
is the most natural for constructing and study of the classical wave
limit of this theory. It is easy to understand, that one must
incorporate all dispersion characteristics to consider both the
corpuscular and wave classical regimes in framework of the single
consistent theoretical scheme. In doing so, one must be sure, that
uncertainty relations are `about minimized'\, during the total
dynamical history of the system. In the opposite case, the wave
pocket can lose its local nature, so the classical theory
description become destroyed without any kinematical `control'\,
under this significant process. Thus, the dispersion
characteristics, or correlations, play extremely important role in
the physically well-motivated formulation of the classical theory
originated from the fundamental quantum one.

Then, using the relation between the dispersions and correlations,
it is easy to check, that the uncertainty relation \be\label{unc}
{\cal D}_{q_m}{\cal D}_{p_m}\geq \frac{\hbar^2}{4} \ee is equivalent
to the inequality \be \left (q_m\circ q_n\right )\,\left (p_m\circ
p_n\right )\geq \hbar^2 \ee written in terms of the correlations.
Our statement is that the inequality (\ref{unc}) remains conserved
under the classical limit procedure defined above. Actually, using
induction in respect to $N$, it is not difficult to prove, that \be
\left ({\cal A}_2^{-1}\right )_{mm}\left ({\cal A}_2+{\cal A}_1{\cal
A}_2^{-1}{\cal A}_1\right )_{mm}\geq 1 \ee for any matrix ${\cal
A}_1$ and for any positively-defined matrix ${\cal A}_2$. This
means, that the uncertainty relations can be incorporated correctly
into the structure of the `soft'\, classical theory on the pure
kinematical level.

Another important quantum inequality between the dispersion
characteristics of the system fixes a signature of the
`quasi-metric',\, which can be related to the total set of the
fundamental correlations. It is given by the relation \be \left
(q_m\circ q_m\right )\,\left (p_m\circ p_m\right )\geq \left
(q_m\circ p_m\right )^2, \ee also remains true in the `soft'\,
classical limit of the theory under consideration. This quantum
relation, which has the explicit Cauchy-Bouniakowsky type,
transforms into the classical inequality \be \left ({\cal
A}_2^{-1}\right )_{mm}\left ({\cal A}_2+{\cal A}_1{\cal
A}_2^{-1}{\cal A}_1\right )_{mm} \geq \left ( \left ({\cal
A}_2^{-1}{\cal A}_1\right )_{mm}\right )^2,\ee which is actually
true for the arbitrary matrices with the properties supposed. Both
the classical inequalities presented above are important for the
study of the kinematics of the theory under construction. They
clarify the features of the dispersion characteristics of motion of
the classical object in the phase space of the system. To
demonstrate a significance of the correlations in the kinematical
structure of the classical system, one needs in development of some
group formalism related to the correlations. We establish it in the
next section.

\subsection{Symplectic formalism and canonical maps}

\setcounter{equation}{0} First of all, let us unify all the
fundamental correlations of the theory to the following $2N\times
2N$ `correlation matrix'\, ${\cal M}$: \be\label{M}
{\cal M}=\left ( \ba{ccc}q\circ q^T&\,&\hbar^{-1}q\circ p^T\\
\hbar^{-1}p\circ q^T&\,&\hbar^{-2}p\circ p^T\ea \right ). \ee In the
`soft'\, classical limit, the explicit form of this block matrix
reads: \be\label{M_c}
{\cal M}_c=\left ( \ba{ccc}{\cal A}_2^{-1}&\,&{\cal A}_2^{-1}{\cal A}_1\\
{\cal A}_1{\cal A}_2^{-1}&\,&{\cal A}_2+{\cal A}_1{\cal
A}_2^{-1}{\cal A}_1\ea \right ), \ee see Eq. (\ref{correl}). It is
clear, that ${\cal M}_c$ is symmetric, i.e., \be\label{sym} {\cal
M}_c^T={\cal M}_c. \ee The less evident algebraic property of the
correlation matrix can be expressed in terms of the quadratic
relation \be\label{sympl} {\cal M}_c{\cal L}{\cal M}_c={\cal L}, \ee
where \be
{\cal L}=\left ( \ba{ccc}0&\,&-1\\
1&\,&0\ea \right ). \ee This group relation can be easily verified
using straightforward calculations and taking into account of Eq.
(\ref{M_c}). The relations (\ref{sym}) and (\ref{sympl}) mean, that
the matrix ${\cal M}_c$ parameterizes the coset $Sp(2N, R) / U(N)$;
in fact it is the canonical null-curvature matrix of the coset
written. Also, it is important to note, that the transformation
\be\label{map} {\cal M}_c\rightarrow {\cal C}{\cal M}_c{\cal C}^T
\ee preserves both the defining coset relations (\ref{sym}) and
(\ref{sympl}), if \be\label{C} {\cal C}{\cal L}{\cal C}^T={\cal L}.
\ee Thus, this transformation does not change algebraic structure
(\ref{M_c}) of the dispersion matrix. This transformation seems like
a symmetry map in the theory of nonlinear sigma-models, and
clarification of its status in the theory under consideration can
generate a real progress in its formulation and in the following
study.

To realize the corresponding program, let us note, that the relation
\be {\cal M}={\cal X}\circ {\cal X}^T,\ee where \be{\cal X}=\left (
\ba {c}q\\p\ea\right )\!,\ee takes place, as it becomes clear
immediately in view of the form of the quantum correlation matrix
given by Eq. (\ref{M}). Thus, in the case of the $q$-independent
transformation matrix ${\cal C}$, one deals with the map
\be\label{X}{\cal X}\rightarrow {\cal C}{\cal X}\ee of the column of
the canonical variables. We stand, that this map, where ${\cal
C}={\cal C}(t_{\alpha})\in Sp(2N, \, R)$, is, in fact, the canonical
transformation in conventional mechanical sense (see
\cite{Str1}--\cite{Str2} for close analogies in superstring gravity
models).

Actually, it is easy to prove, that \be\left \{{\cal X},\, {\cal
X}^T\right \}={\cal L},\ee where it is taken into account, that the
Poisson bracket between the moment $p_m$ and coordinate $q_n$ is $\{
p_m,\,q_n\}=\delta_{mn}$. Thus, the group property (\ref{C}) for the
transformation ${\cal C}$ is equal to the usual conservation of the
fundamental Poisson brackets. This proves our statement: the ${\cal
C}$-map is actually the conventional canonical transformation.

Then, the symplectic quantities established allow one to write down
the `soft'\, Hamilton equations for the system under consideration
in their explicitly symplectic invariant form. Here we mean
generalization of the special Hamilton equations, written for the
`hard'\, regime with $\hbar=0$. Namely, the statement is: in the
`soft'\, case under discussion with $\tilde {\cal H}_{\alpha}=\tilde
{\cal H}_{\alpha}(q,\, p)=\tilde {\cal H}_{\alpha}({\cal X})$, the
leading symplectic term (which does not include $\hbar$) in the
motion equation of the theory reads: \be\label{Ham} {\cal
X}_{,\,t_{\alpha}}\cong-{\cal L}\nabla_{{\cal X}} \tilde {\cal
H}_{\alpha}+\epsilon_{\alpha\beta}{\cal M}_c\nabla_{{\cal X}} \tilde
{\cal H}_{\beta},\ee where
$\epsilon_{\alpha\beta}=-\epsilon_{\beta\alpha}, \,\,
\epsilon_{12}=1$ is the Levi-Chivita symbol. Here $\tau=t_1-it_2$
and ${\cal H}={\cal H}_1+i{\cal H}_2$ now (let us remember, that
$\hbar\neq 0$ in the `soft'\, classical regime under consideration).
Then, in (\ref{Ham}) the total derivative in respect to ${\cal X}$
is understood as the $2N$-column. Also we mean, that the quantities
$\tilde {\cal H}_{\alpha}$ are considered as the functions
calculated in the classical limit given by substitution of
$\nabla_{q}{\cal S}_1$ to the $p$-term into the Hamiltonians. The
exact form of the right side of Eq. (\ref{Ham}) contains terms
depending on the Plank constant. It is interesting to note that the
underlying symplectic structure is totaly $\hbar$-free. Note, that
one has a reason to consider this system of equations in the case
with finite values of $t_2$ and ${\cal H}_2$, i.e., when
$\hbar\beta$ and $\hbar\gamma$ are non-vanishing quantities. Thus,
the corresponding `soft'\, limit must be `viewed'\, at high
temperatures and very fast decay processes in the system.

Also, one must answer the natural question, when it is possible to
neglect the $\hbar$-depending terms in the modified Hamiltonian
equation (\ref{Ham}). It is not difficult to prove, that in the
corresponding case the consistency condition ${\cal
X}_{,\,t_1t_2}={\cal X}_{,\,t_2t_1}$ is satisfied if \be\label{eqM}
{\cal M}_{,\,t_{\alpha}}={\cal M}I_{\alpha}{\cal L}-{\cal
L}I_{\alpha}{\cal M}+\epsilon_{\alpha\beta}\left ({\cal
L}I_{\beta}{\cal L}+{\cal M}I_{\beta}{\cal M}\right ),\ee where
$I_{alpha}=\nabla_{\cal X}\nabla^T_{\cal X}{\cal H}_{\alpha}$.
Finally, Eq. (\ref{eqM}) is consistent itself, if the Hamiltonians
${\cal H}_{\alpha}$ are some linear or quadratic functions in
respect to the canonical variables ${\cal X}$. Note, that these
variables can be related to the original physical coordinates and
moments in a highly nontrivial form, so the last restriction is not
too hard, as it can be naively understood. For example, one can use
the action-angle variables, which can be taken in the corresponding
form for any Hamiltonian system of the type under consideration
(i.e., one can reach linear or quadratic form of the Hamiltonian
functions in the general case).

Then, using straightforward calculations, it is not difficult to
prove, that in the isothermal and adiabatical regimes of the
thermodynamical evolution, this new (`soft') classical system
demonstrates the effect of the irreversibility in its evolution. The
proof is very similar to the one performed in the previous section;
we will leave it as the not really difficult exercise. It is
important to stress, that this leading term in the modified Hamilton
equations (\ref{Ham}) is actually invariant under the transformation
(\ref{M}), (\ref{X}).

For study and application of the canonical map established above,
let us introduce the following complex symmetric $N\times N$-matrix
field $\cal A$: \be {\cal A}={\cal A}_1+i{\cal A}_2. \ee Our
statement is that the transformation (\ref{map}) can be decomposed
into set of the maps \be\label{set} {\cal A}\rightarrow {\cal
A}+\Lambda_1,\quad {\cal A}\rightarrow \Lambda_2^T{\cal
A}\Lambda_2,\quad {\cal A}^{-1}\rightarrow  {\cal A}^{-1}+\Lambda_3,
\ee where $\Lambda_1, \Lambda_2$ and $\Lambda_3$ are the real
$t_{\alpha}$-dependent matrices, $\Lambda_1^T=\Lambda_1,\,\,
\Lambda_3^T=\Lambda_3$, and ${\rm det}\Lambda_2\neq 0$. It is
interesting to note, that these transformations coincide exactly
with the matrix-valued hidden symmetries of the General Relativity
written in terms of the Ernst (matrix) potential ${\cal A}$. Namely,
the first map from this set is the shift transformation, the second
one is the scale symmetry, whereas the third map is the Ehlers
non-linear transformation. A remarkable fact related to the
presented decomposition is based on the following trivial statement:
the discrete map \be\label{discr}{\cal A}\rightarrow {\cal
A}^{-1}\ee transforms the shift symmetry into the Ehlers map, and
vice-versa. Also, the scale transformation remains itself under this
discrete map. One can prove, that the map (\ref{discr}) corresponds
to the transformation (\ref{map}) with ${\cal C}={\cal L}$, which
generates the canonical interchange \be q\rightarrow -p, \qquad
q\rightarrow p.\ee Thus, this discrete map is equivalent to the
change of representation in the original quantum theory. This means,
that kinematics of the classical theory theory under consideration
allows the corresponding degree of freedom -- the really promising
and important fact. From this it follows, for example, that our
formalism can be successfully used for description of the both
corpuscular and wave limits of the original quantum theory.

Also it is clear, that change of the representation will be
reflected on the explicit form of the dispersion characteristics of
the system. These dispersion characteristics, or the classical
correlations, define effectively some `tube of motion'\, in the
nearest vicinity of the classical trajectory of the system. They
give a key to the kinematic control under the type of the motion:
namely, a continuous increasing of the `tube diameter'\, makes the
average trajectory of the classical object more and more
`quantum-like'\, one. Also, it seems an obvious fact, that the
matrix ${\cal M}_c$ can be considered as new field, which makes the
phase space of the theory curved in a quasi-gravitational sense.
This matrix defines the analogy of the metric in the phase space,
which `interacts'\, with the `matter degrees of freedom',\,
collected in the canonical variable ${\cal X}$.

At the end of this section let us stress, that the transformation
set (\ref{set}) can be used for the really dramatic simplification
of the dynamical problem for this classical system. The
corresponding details for the arbitrary symplectic coset one can
find in the literature. The statement is that one can use these
transformations to move the ${\cal A}_{\alpha}$-values to the their
simplest possible form ${\cal A}_1=0, \, {\cal A}_2=1$. Note, that
this simplification can be performed during the whole dynamical
history of the system (in the any given point of its classical
trajectory). One can use this nontrivial fact to perform the
dynamical analysis of the complete classical theory in the most
simple form.

\subsection{Conclusion}

\setcounter{equation}{0}

In this part we have developed both the kinematical and dynamical
parts of the classical theory, which arises in the consistent limits
of the quantum theory with arrow of time. we have studied `hard'\,
and `soft'\, classical regimes of the theory, which were defined as
the ones describing resonances of the probability density, with
$\hbar=0$ and $\hbar\neq 0$, respectively. We have shown, that it is
natural to add all the dispersion characteristics (or the
correlations) to the `usual'\, set of the fundamental dynamical
variables -- i.e., to the canonical coordinates and moments of the
classical theory. In particular, it is established, that in the
`soft'\, regime the resulting classical theory possesses a
symplectic group of the conventional canonical transformations. It
is proved the fulfilment of the uncertainty relations in the
presented classical kinematics. We have proposed the
quasi-gravitational interpretation of the dispersion parameters and
conception of the curved phase space.

We have shown, that the modified classical dynamics allows a
well-defined arrow of time (at least for the isothermal and
adiabatic regimes). In the `hard'\, classical regime, we have
studied in details evolution of the classical limit of the original
quantum system using the most convenient representation, which
corresponds to the action-angle one in the standard classical
mechanics. We have derived the modified Hamiltonian equations and
prove their consistency condition in the case of the general
canonical representation taken.

In the `soft'\, representation case, we have established the compact
matrix form of the modified Hamiltonian equations which is
explicitly invariant under the action of the hidden subgroup of
canonical symmetries. We have studied and classified all these
symmetries and discover their quasi-Einsteinium structure. In
particular, we have extracted from this symmetry subgroup the
nonlinear sector of Ehlers transformations. Its arising (as well as
the metric-like behavior of the classical correlations in the
theory) demonstrates real possibilities for the following
development of this theory in the General Relativity direction. It
seems natural to wait different applications of the formalism
developed in all gravity involved theories -- string theory,
cosmology, black hole physics \cite{PartPhysCosm-f}--\cite{Str-f}.
We hope to present corresponding results in the forthcoming works.

\vskip 10mm \noindent {\large \bf Acknowledgements}

\vskip 3mm \noindent We would like to thank prof. B.S. Ishkhanov for
his continuous help and attention to us and this work during all
time of its preparation. In fact, many of general and special
problems considered and solved in this publication were inspired by
our numerous discussions and private talks. We think, that they were
obtained in many respects owing to this remarkable human and
scientific collaboration.

\end{document}